\documentclass{article}

\usepackage{amsmath,amssymb,amsfonts,dcolumn,color,graphicx,graphics,latexsym,placeins,epsfig}
\usepackage{graphicx}  
\usepackage{amsmath}   
\usepackage{amssymb}   
\usepackage{bm} 
\usepackage{dcolumn}
\usepackage{color}
\usepackage{footmisc}
\usepackage{mathrsfs}
\usepackage{amsfonts}
\usepackage{varioref}
\usepackage{slashed}
\usepackage{footnote}
\usepackage[belowskip=-17pt,aboveskip=15pt]{caption}
\usepackage{amsmath}
\usepackage{makeidx}
\usepackage{amssymb}
\usepackage{graphicx}
\usepackage{bmpsize}
\usepackage{multicol}
\usepackage{url}
\usepackage{ragged2e}

\usepackage[margin=1.5in]{geometry}
\usepackage{graphicx}
\usepackage{dcolumn}
\usepackage{amsmath,amssymb,mathtools}
\usepackage{epstopdf}
\usepackage{verbatim}
\usepackage[colorlinks=true, citecolor=blue, urlcolor = blue, linkcolor= blue, 
bookmarks=true]{hyperref}
\usepackage[utf8]{inputenc}
\usepackage{amsmath}
\def\e{\epsilon}
\def\p{\partial}

\def\be{\begin{eqnarray}}
\def\ee{\end{eqnarray}}

\def\e{\epsilon}
\def\p{\partial}
\labelformat{section}{Section #1} 
\labelformat{subsection}{Section #1} 
\labelformat{subsubsection}{Section #1}
\labelformat{subsubsubsection}{Section #1}
\labelformat{eqnarray}{Eq.~(#1)} 
\labelformat{figure}{Fig.~#1} 
\labelformat{subfigure}{Fig.~\thefigure#1} 
\labelformat{table}{Tab.~#1} 
\labelformat{appendix}{Appendix #1}
\vspace{25pt}
\title{\bf $\xi R\phi^2$ non-minimal coupling, and the  long range gravitational potential for different spin fields from 2-2 scattering amplitudes}
\author{\bf Avijit Sen Majumder\footnote{senmajumderavijit@gmail.com}, \,\,  Ayan Kumar Naskar\footnote{meayannaskar@gmail.com} \,\, and \,\,  Sourav Bhattacharya\footnote{sbhatta.physics@jadavpuruniversity.in}\\
\small{Relativity and Cosmology Research Centre, Department of Physics, Jadavpur University, Kolkata 700 032, India}\\}

\begin{document}
\maketitle

\begin{abstract}
\noindent
In this paper, we investigate the long range gravitational effect of  curvature-scalar field non-minimal coupling in the form of $\xi R \phi^2$, in the  perturbative quantum gravity framework.  Such coupling is most naturally motivated from the renormalisation of a scalar field theory with a quartic self interaction in a curved spacetime background. This coupling results in two scalar-$n$ graviton vertices which contain no explicit momenta of the scalar, qualitatively different from the usual, e.g., $\kappa h^{\mu\nu}T_{\mu\nu}$-type minimal matter-graviton vertices. Assuming the dimensionless  coupling parameter $\xi$ to be small, we compute the 2-2 scattering Feynman amplitudes between such scalars up to ${\cal O}(G^2 \xi)$. From the non-relativistic limit of these amplitudes, we compute the corresponding long range gravitational potential. There exists no tree level contribution $({\cal O}(\xi G))$ here, and hence the one loop ${\cal O}(G^2 \xi)$ result is leading. Recently, the effect of a cosmological constant in such non-minimal interaction and the subsequent gravitational potential was computed. In this work, we take the cosmological constant to be vanishing. The resulting potential is found to have $r^{-4}$ leading behaviour. We further extend these results for scalar-massive spin-1 and massive spin-1/2 scattering. Spin and polarisation dependence of the two body potential have been explicitly demonstrated. We discuss some possible physical implications of these results.
\end{abstract}

\section{Introduction}\label{S1}

Among the four fundamental interactions of nature, i.e., strong, weak, electromagnetic, and gravity, the first three can be quantised, and the resulting quantum field theories make predictions that are in excellent agreement with observation.  A traditional approach to quantising general relativity, however, shows that it is not perturbatively renormalisable. At each order of expansion of the spacetime inverse metric with respect to a classical background, we generate new terms in the action, and accordingly we need new counterterms at every order of perturbation theory for renormalisation. This was established in a series of pioneering works such as~\cite{Utiyama, Weinberg:1965nx, Pagels:1966zza, vanDam:1970vg, Zakharov:1970cc, Iwasaki:1971vb, Boulware:1972yco, tHooft:1974toh, deser1, deser2, deser3, Stelle, Voronov, Goroff:1985th,  Odintsov4, Odintsov5, PRD:87, Lavrov}.  The fact that gravity is not a renormalisable quantum field theory has led to many alternative ideas for a consistent quantum theory of gravity. However, it will be fair enough to admit that none of them has given us any complete or satisfactory answer so far, see e.g.~\cite{Mukhanov, Keifer, Shapiro}, and references therein. 

Despite this shortcoming or incompleteness of gravity while treated as a quantum field theory, it is believed that at energy scales much below the Planck mass, at least the first couple of orders of computation in this framework might be physically meaningful. The predictions thus made can be hoped to be tested in the not too far away future, especially keeping in mind the improving observational capacity to probe strong gravity regime. Such observations can tell us whether there is any  deviation from the classical theory of gravity. In particular, they can tell us whether the notion of graviton as the quantum of gravity makes any sense.  Second, one also hopes that these low energy computations will one day be successfully embedded in a more complete theory of quantum gravity. With this motivation, there has  really been a huge amount of effort by the community over past few decades to understand the physical predictions of perturbative quantum gravity, see e.g.~\cite{Hiida:1972xs, Barker:1975ae, Donoghue:1993eb, Donoghue:1994dn, Muzinich:1995uj, SD1, SD2, Hamber:1995cq, MODANESE1995697, Akhundov:1996jd, Donoghue_1996, Bjerrum-Bohr:2002aqa, Bjerrum-Bohr:2002fji, Bjerrum-Bohr:2002gqz, Bern:2002kj, Akhundov:2006gh, Holstein:2008sx,Toms:2010vy, Bjerrum-Bohr:2015vda, Malta2015ComparativeAO, Frob:2016xte,Ulhoa_2017, Olyaei:2018asy, deBrito:2020wmp, Frob:2021mpb, Majumder:2025sht, Donoghue:2017pgk} and also references therein. They chiefly  investigate  the quantum corrections of the long range, two body gravitational potentials, renormalisation of matter-graviton interaction processes, and running of various couplings. See also~\cite{Bjerrum-Bohr:2014zsa, Bai:2016ivl, Bastianelli:2021nbs} for computation of gravitational light bending at leading and subleading orders using graviton exchanges between massive and massless fields. We  refer our reader to~\cite{Elizalde, Toms:2008dq, Toms:2009zz, Toms:2009vd, Toms:2011zza, Majumder:2025gou} for inclusion of a cosmological constant in the perturbative quantum gravity calculations, for processes occurring much inside the cosmological event horizon. We further refer our reader to e.g.~\cite{Donoghue:1995cz, Burgess:2003jk,Goldberger:2004jt, Foffa:2016rgu, Levi:2018nxp, Cheung:2018wkq, Bern:2020buy, Ivanov:2022qqt, Almeida:2024uph, Almeida:2024cqz, Trestini:2024mfs} (also references therein), for an effective field theory description of quantum gravity. These computations also do not concern with the ultraviolet completion of  gravity, and chiefly focus on  the two body long range gravitational potential at various post Minkowski orders, as well as on gravitational radiation. Finally, see e.g.~\cite{Odintsov2, Muta:1991mw, SD3, SD4, Seery:2007we, Woodard:2014jba, Oda:2015sma, Moss:2014nya, Shapiro:2015ova, Saltas:2015vsc, Buchbinder, Arbuzov:2021yai} and references therein for discussion on perturbative quantum gravity in curved, especially the primordial inflationary de Sitter, backgrounds.

When one tries to renormalise a scalar field theory with a quartic self interaction in the presence of gravity,   inclusion of a matter-gravity non-minimal term like $\xi R \phi^2/2$ in the action becomes mandatory, see~\cite{Parker:2009uva} and the original references therein. Note also that $\xi=1/6$ in particular leads to the trace or conformal anomaly in the massless limit. With a conformal scalar field, stationary hairy back hole solutions can be seen in~\cite{Bocharova:1970skc, Bekenstein:1974sf, Martinez:2002ru, Bhattacharya:2013hvm, Bravo-Gaete:2025vyd}.   What will be the contribution of such non-minimal interaction to the two body long range gravitational potential?  While the minimal  case has been extensively investigated in the literature, e.g.~\cite{Hamber:1995cq, Bjerrum-Bohr:2002aqa, Bjerrum-Bohr:2002fji}, (also e.g.~\cite{Cheung:2018wkq} and references therein for effective field theory computations), the case $\xi \neq 0$ seems to be  relatively less addressed.  This coupling generates  two scalar-$n$ graviton vertices, each containing explicitly the momenta carried by the graviton lines, but  not of the two scalar lines,  somehow complementing the usual matter-graviton vertices (like $\kappa h_{\mu\nu} T^{\mu\nu}$), where only the momenta carried by the matter field appear explicitly.  
 
In this paper we wish to compute the effect of such non-minimal coupling in the two body gravitational potential, via the computation of  the 2-2  scattering amplitudes between massive scalar-massive scalar, massive scalar-massive spin-1, and massive scalar-massive spin-1/2 fields.  To the best of our knowledge, this question was first addressed in~\cite{Majumder:2025gou}, where the effect of the three graviton vertex generated by the cosmological constant (via the $-2\Lambda \sqrt{-g}/\kappa^2$ term in the action) was studied for scalar-scalar scattering. In this paper, we will set $\Lambda=0$. The rest of the paper is organised as follows. In the next section, we briefly discuss the basic ingredients we will be needing for our computations. In \ref{scalar}, we compute the non-minimal  gravitational scattering between two massive scalar fields. In \ref{proca} and \ref{dirac}, we respectively compute the same for massive scalar-massive spin-1 and massive scalar-massive spin-1/2 fields. The leading behaviour of the potentials turns out to be $\sim r^{-4}$. We compare our result with the well known $\xi=0$ case in  \ref{disc}. The non-minimal interaction will be taken only for the scalar field.  We will assume in the following that the coupling parameter $\xi$ is small, so that we will restrict our computations to linear order in $\xi$ only.\\
\noindent
We will work with the mostly positive signature of the metric in four spacetime dimensions. The incoming (outgoing) momenta in the 2-2 scattering process will always be denoted by $k_1$, $k_2$ $(k'_1 , k'_2)$ respectively, so that  $k_1+k_2= k'_1+k'_2$. $k_1$ and $k'_1$ will be associated with the  scalar, whereas $k_2, k'_2$ will stand for various massive spin fields (0,1,1/2). All these external momenta  will  be taken to be non-relativistic. The transfer momentum $(k_1 - k'_1)=(k'_2 - k_2)$ will be denoted by $q$. For symmetrisation, we will use the notation in this paper : $X_{(\alpha\beta)} = X_{\alpha} Y_{\beta}+X_{\beta} Y_{\alpha}$. 

\section{The basic ingredients}
Let us first briefly discuss the basic ingredients we will be needing for our main computations. Much of the discussion  appearing below can be seen in e.g.~\cite{Holstein:2008sx} (see also references therein). We begin with the action of the theory,
\begin{eqnarray}
    \begin{split}
        S=& \frac{2}{\kappa^2}\int d^4 x \sqrt{-g} R  - \frac12 \int d^4 x \sqrt{-g}\bigg[g^{\mu\nu}(\nabla_{\mu}\phi) (\nabla_{\nu}\phi)+  (M^2 + \xi R) \phi^2\bigg] -\frac12 \int d^4 x \sqrt{-g}\bigg[g^{\mu\nu}(\nabla_{\mu}\varphi) (\nabla_{\nu}\varphi)  \\ 
& + (m^2+\xi R) \varphi^2 \bigg]
+\int d^4 x \sqrt{-g} \bigg(i\bar{\Psi} \slashed{\nabla} \Psi -m_f \bar{\Psi}\Psi \bigg) - \int d^4 x \sqrt{-g} \bigg(\frac14g^{\mu\rho}g^{\nu\lambda} F_{\mu\nu}F_{\rho\lambda} + \frac12 m_v^2 g^{\mu\nu} A_{\mu} A_{\nu} \bigg),
\label{nmadd1}
    \end{split}
\end{eqnarray}
where $\xi$ is the dimensionless non-minimal coupling parameter, $\kappa^2= 32\pi G$, and $F_{\mu\nu}= \p_{\mu}A_{\nu}-\p_{\nu}A_{\mu}$. $\slashed{\nabla}\equiv \tilde{\gamma}^{\mu}\nabla_{\mu}$, where $\nabla$ acting on the spinor is the spin covariant derivative. Also, since we are working with the mostly positive signature of the metric, the anti-commutation relation for the curved space $\gamma$-matrices reads
$$[\tilde{\gamma}^{\mu}, \tilde{\gamma}^{\nu}]_+ = -2 g^{\mu\nu} {\bf I}.$$
We wish to compute the scattering of $\varphi$,  $A_{\mu}$ and $\Psi$ off the scalar $\phi$, one by one. We will assume $\xi$ to be small, and will compute the amplitudes up to ${\cal O}(\xi G^2)$. Note also that we have  kept the same value for the non-minimal coupling for the second scalar,  $\varphi$. Extension with two different values of them is straightforward,  although a little bit more tedious. \\

\noindent
We will work in the weak gravity regime where the background can be taken to be Minkowski, so that  we have the series expansion
\begin{eqnarray}
 &&g_{\mu\nu}= \eta_{\mu\nu}+\kappa h_{\mu\nu}, \qquad \qquad g^{\mu\nu}= \eta^{\mu\nu}-\kappa h^{\mu\nu}+ \kappa^2 h^{\mu\alpha} h_{\alpha}{}^{\nu} -\kappa^3h^{\mu}{}_{\alpha} h^{\alpha}{}_{\beta}h^{\beta\nu} +\cdots, \nonumber \\ &&\qquad \sqrt{-g}= 1+ \dfrac{\kappa h}{2} + 
 \dfrac{\kappa^2}{4} \left(\dfrac{ h^2}{2}-  h_{\mu\nu}h^{\mu\nu}\right) +\dfrac{\kappa^3}{48}\left(h^3-6h\,h_{\mu\nu}h^{\mu\nu}
+8h^\mu{}_\nu h^\nu{}_\rho h^\rho{}_\mu
\right)+\cdots \nonumber \\ &&
\Gamma^{\mu}_{\nu\rho} = \dfrac{\kappa}{2} \eta^{\mu\alpha} \left(\p_{\nu} h_{\rho\alpha}+\p_{\rho} h_{\nu\alpha} - \p_{\alpha} h_{\nu\rho}\right) -\dfrac{\kappa^2}{2} h^{\mu\alpha} \left(\p_{\nu} h_{\rho\alpha}+\p_{\rho} h_{\nu\alpha} - \p_{\alpha} h_{\nu\rho}\right)
+ \dfrac{\kappa^3}{2} h^{\mu\beta} h_{\beta}{}^{\alpha} \left(\p_{\nu} h_{\rho\alpha}+\p_{\rho} h_{\nu\alpha} - \p_{\alpha} h_{\nu\rho}\right)
+\cdots \nonumber\\
\label{nmadd2}
\end{eqnarray}
Only up to ${\cal O}(\kappa^2)$ expansion of various quantities appearing above  will effectively be necessary for our present tree (${\cal O}(\kappa^2)$) and one loop (${\cal O}(\kappa^4)$) level computations, as follows. We need 2-matter-1-graviton, 2-matter-2-graviton, and also the 3-graviton vertices. The first and the second require the expansion of the inverse metric and $\sqrt{-g}$ up to ${\cal O}(\kappa^2)$, for the minimal case. For the third, on the other hand, one needs to expand $\sqrt{-g}R=\sqrt{-g}g^{\mu\nu}R_{\mu\nu}$  at ${\cal O}(\kappa^3)$.  However, the $\kappa^2$ sitting in the denominator of this term  in the Einstein-Hilbert action (Eq.~\ref{nmadd1}) effectively reduces it to  ${\cal O}(\kappa)$. Now, since the lowest order expansion of the metric and its determinant is ${\cal O}(\kappa^0)$, we must expand $R_{\mu\nu}$ up to ${\cal O}(\kappa^3)$. This means that for the terms in the Ricci tensor containing a single $\Gamma$, we must take $\Gamma$  up to ${\cal O}(\kappa^3)$. However,  the $\Gamma(\kappa^3)$ terms yield total divergences, and hence they do not contribute to the three-graviton interaction. Note also that for the non-minimal term $\xi \sqrt{-g} R\phi^2$, there is no $\kappa^2$ sitting in the denominator, Eq.~\ref{nmadd1}, and hence this term requires an expansion of $\sqrt{-g} R$ only up to  ${\cal O}(\kappa^2)$. This will be more explicit from the following calculations.\\

\noindent
In the de Donder gauge,
$$\p_{\mu}\left(h^{\mu}{}_{\nu}-\frac12 \delta^{\mu}_{\nu} h \right)=0,$$
the graviton propagator reads
\begin{eqnarray}
&& \Delta_{\mu\nu\alpha\beta}(k)= -\frac{i{\cal P}_{\mu\nu\alpha\beta}}{k^2},
\label{nmadd3}
\end{eqnarray}
where 
\begin{eqnarray}
&& {\cal P}_{\mu\nu\alpha\beta}= \frac12 \left( \eta^{\mu\alpha} \eta^{\nu\beta}+\eta^{\mu\beta} \eta^{\nu\alpha} - \eta^{\mu\nu} \eta^{\alpha\beta}\right).
\label{nmadd4}
\end{eqnarray}
The tree level scalar  propagator with mass $m$ reads 
\be
\Delta (k) = -\frac{i}{k^2+m^2}.
\ee

\noindent
Let us now come to various vertex functions we will require for our purpose.  By setting $\xi =0 $ in Eq.~\ref{nmadd1},  the two scalar-one graviton minimal vertex reads, e.g.~\cite{Bjerrum-Bohr:2002fji}, 
\begin{eqnarray}
V_{\text{spin-0}}^{(1)\,\mu \nu} (k,k',m)= -\frac{i\kappa}{2} \Big[k^{\mu}k'^{\nu}+k'^{\mu}k^{\nu}-\eta^{\mu\nu}\Big(k\cdot k' +m^2 \Big) \Big],
\label{qg5'}
\end{eqnarray}
whereas the two scalar-two graviton minimal vertex reads,
\begin{eqnarray}
    \begin{split}
    \label{qg21}
        V_{\text{\:spin-0}}^{(2)\,\mu \lambda \rho \sigma} (k,k',m) =& i\kappa^2 \Big[\Big\{ I^{\mu \lambda \alpha \nu}  I^{\rho\sigma \beta}{}_{\nu} -\frac14 \big(\eta^{\mu \lambda}I^{\rho\sigma \alpha \beta} +\eta^{\rho \sigma}I^{\mu \lambda \alpha \beta}  \big)  \Big\} \Big(k_{\alpha}k'_{\beta}+k_{\beta}k'_{\alpha} \Big)\\ 
        & -\frac12 \Big(I^{\mu \lambda \rho \sigma} -\frac12 \eta^{\mu \lambda}\eta^{\rho\sigma} \Big) \Big(k\cdot k' +m^2 \Big) \Big], 
    \end{split}
\end{eqnarray}
where, 
$$I_{\mu\nu\lambda \rho}= \dfrac12 \Big( \eta_{\mu\lambda} \eta_{\nu\rho} + \eta_{\mu\rho} \eta_{\nu\lambda}\Big).$$
Note that the momenta appearing in Eqs.~\ref{qg5'}, \ref{qg21} are carried by the two scalar lines, originating from the derivative acting on the scalar. The non-minimal vertex, however, will  just be the  opposite of this, as will be demonstrated towards the end of this section.
 \\

\noindent
For the spin-1 field with mass $m_v$, the propagator reads (e.g.~\cite{Coleman:2018mew})
\begin{eqnarray}
    D_{\mu \nu}(k) = -\frac{i}{k^2 +m_v^2} \Bigg(- \eta_{\mu \nu} + \frac{k_{\mu} k_{\nu}}{m_v^2} \Bigg).
\end{eqnarray}
 The two massive spin-1-one graviton and the two massive spin-1-two graviton vertices respectively read \cite{Holstein:2008sx}
\begin{eqnarray}
    \begin{split}
         V^{\text{spin-1}\,(1)}_{\beta,\alpha;\mu\nu}(k,k',m_v) =& - {i \kappa \over 2}   \eta_{\mu \nu} \Big[(k \cdot k'+ m_v^2) \eta_{\alpha \beta} - k_{(\alpha} k'_{\beta)} \Big] + i \kappa   I_{\mu \nu \kappa \lambda} \Big[(k \cdot k'+ m_v^2) {I_{\alpha \beta}}^{\kappa \lambda}   \\
         & + \frac{1}{2} k^{(\kappa} k'^{\lambda)}  \eta_{\alpha \beta} - \big(k^\kappa k'_{\alpha} \delta_\beta^\lambda + k'^\kappa k_{\beta} \delta_\alpha^\lambda \big)\Big], \\
    \end{split}
\end{eqnarray}
and,
\begin{eqnarray}
    \begin{split}
         V^{\text{spin-1}\,(2)}_{\beta,\alpha,\mu\nu \rho \sigma}(k,k', m_v) =&  {i \kappa^2 \over 2}  {\cal P}_{\mu \nu \rho \sigma} \Big[(k \cdot k'+ m_v^2) \eta_{\alpha \beta} - k_{\beta} k'_{\alpha} \Big]  -  i \kappa^2 \Big\{  {I_{\mu \nu}}^{\kappa \delta} {I_{\rho \sigma \delta}}^{\lambda} +  {I_{\rho \sigma}}^{\kappa \delta} {I_{\mu \nu \delta}}^{\lambda} -  \frac{1}{2} \Big(\eta_{\mu \nu} {I_{\rho \sigma}}^{\kappa \lambda} \\
         &  +  \eta_{\rho \sigma} {I_{\mu \nu}}^{\kappa \lambda}\Big)  \Big\}  \Big[(k \cdot k'+ m_v^2) I_{\alpha \beta\kappa \lambda} + \frac{1}{2} \big(k_{\kappa} k'_{\lambda} + k_{\lambda} k'_{\kappa}\big) \eta_{\alpha \beta} - \big(k_{\kappa} k_{\alpha} \delta_{\beta \lambda} + k'_{\kappa} k_{\beta} \delta_{\alpha \lambda} \big)\Big] \\
         &  - \frac{i \kappa^2 }{2}  \Big({I_{\mu \nu}}^{\eta \theta} {I_{\rho \sigma}}^{\kappa \lambda} + {I_{\rho \sigma}}^{\eta \theta} {I_{\mu \nu}}^{\kappa \lambda} \Big)  
 \Big[k_{\eta} \eta_{\alpha \kappa} (k'_{\theta} \eta_{\beta \lambda} -  k'_{\lambda} \eta_{\beta \theta} ) +  k'_{\eta} \eta_{\beta \kappa} (k_{\theta} \eta_{\alpha \lambda}  -  k_{\lambda} \eta_{\alpha \theta} )  \Big],
    \end{split}
\end{eqnarray}
where  $I_{\mu \nu \kappa \lambda}$ can be seen below Eq.~\ref{qg21}. 
Let us now clarify the issue of the polarisation, say $\e^{\mu}(k)$, of the massive spin-1 field. Since the field has three independent degrees of freedom, we have the constraint $k\cdot \e=0$. Since the transfer momentum is given by ${k}_1- {k}'_1= {k}'_2-{k}_2=q$,  we first write
\begin{equation}
\vec{k}_1= \vec{k} + \frac{\vec q}{2}, \qquad \vec{k}'_1= \vec{k} - \frac{\vec q}{2}; \qquad \qquad \vec{k}_2= \vec{k} - \frac{\vec q}{2}, \qquad \vec{k}'_2= \vec{k} + \frac{\vec q}{2},
\label{qgs2}
\end{equation}
where $\vec{k}$ is the momentum of the centre of mass frame. We next write  in the non-relativistic limit for the polarisation vector~\cite{Holstein:2008sx},
\begin{equation}
 \epsilon^{\mu}(k)\vert_{\rm NR} \simeq \left( \frac{\vec{k}\cdot \vec{\epsilon}}{m_v},\ \vec{\epsilon}\right) 
 \label{pol}
 \end{equation}
which trivially satisfies $k\cdot \epsilon=0$ for $k_0 \approx -m_v$. We have
\begin{equation}
\epsilon_{\mu}(k_1)\,\epsilon^{\,\star \, \mu}(k'_1)\vert_{\rm NR}= \vec{\epsilon}\,(\vec{k}_1)\cdot \vec{\epsilon}^{\,\,\star}(\vec{k}'_1)- \frac{i}{2m_v^2}\,\vec{S} \cdot (\vec{k}\times \vec{q})-\frac{1}{m_v^2} \, \vec{k}\cdot \vec{\epsilon}\,(\vec{k}_1)\ \vec{k}\cdot \vec{\epsilon}^{\,\,\star}(\vec{k}'_1) +\frac{1}{4m_v^2} \, \vec{q}\cdot \vec{\epsilon}\,(\vec{k}_1)  \ \vec{q}\cdot \vec{\epsilon}^{\,\,\star}(\vec{k}'_1),
\label{qgs3}
\end{equation}
where we have written for the spin vector, 
\begin{equation}
    -i\vec{S} = \vec{\epsilon}\,(\vec{k}_1)\times \vec{\epsilon}^{\,\,\star}(\vec{k}'_1),
\label{qgs3'}
\end{equation}
and have  used the trivial identity,
$$\vec{q}\cdot \vec{\epsilon} \ \vec{k}\cdot \vec{\epsilon}\,\,' = \vec{k}\cdot \vec{\epsilon} \ \vec{q}\cdot \vec{\epsilon}\,\,'+ i\vec{S}\cdot (\vec{k}\times \vec{q}).$$
We use the abbreviation in the above and below : $\vec{\e}^{\,\,\star}(k')\equiv \vec{\e}\,\,'$, for the sake of conveneience.\\

\noindent
Let us next come to the case of the massive spin-1/2 field. The propagator reads
\begin{eqnarray}
    \begin{split}
        S (k) = & -\frac{i (\slashed{k} + m_f)}{k^2 +m_f^2}.
    \end{split}
    \label{fprop}
\end{eqnarray}
The relevant three and four point vertices respectively read \cite{Holstein:2008sx}
\begin{eqnarray}
    \begin{split}
        V^{\text{spin-1/2}\,(1)}_{\mu\nu} (k, k', m_f) & = -{i\kappa\over 2}\Bigg[{1\over 8}\gamma_{(\mu}(k  +  k')_{\nu)}  -  \eta_{\mu\nu}  \Big({1\over 2}(\! \not\!{k} +  \!\not\!{k'}) -  m_f  \Big) \Bigg],
    \end{split}
\end{eqnarray}
and
\begin{eqnarray}
    \begin{split}
      V^{\text{spin-1/2}\,(2)}_{\mu\nu\rho\sigma}&(k, k', m_f)  =  i\kappa^2\Big[ -{1\over 2} \bigg \{{1\over 2}(\not \!{k}  + \! \not\!{k'} ) - m_f \bigg\}  {\cal P}_{\mu\nu\rho\sigma} - {1\over 32} \bigg\{\eta_{\mu\nu} \gamma_{(\rho}(k+k')_{\sigma)}  +  \eta_{\rho\sigma}\gamma_{(\mu }(k+k')_{\nu)} \bigg\}  \\
     & + {3\over 16}(k+k')^{\epsilon}\gamma^{\xi} (I_{\xi\phi\mu\nu}{I^{\phi}}_{\epsilon\rho\sigma} +I_{\xi\phi\rho\sigma}{I^{\phi}}_{\epsilon\mu\nu}) + {i\over 16}\epsilon^{\epsilon\phi\eta\lambda}\gamma_\lambda \gamma_5 \Big(I_{\rho\sigma\phi\xi} {I_{\mu\nu\eta}}^\xi  {k}_\epsilon - I_{\mu\nu\phi\xi} {I_{\rho\sigma\eta}}^\xi  (k+q)_\epsilon \Big)\Big], 
    \end{split}
\end{eqnarray}
where in both the expressions above, the momenta are carried by the spin-1/2 field. We will also need the Gordon identity satisfied by the spinors, 
\begin{equation}
    \bar{u}_{s'}(k'_2) \gamma^{\mu} u_s(k_2)=\frac{1}{2m_f}\bar{u}_{s'}(k'_2)\left[ (k_2+k'_2)^{\mu}-\frac12 [\gamma^{\mu},  \gamma^{\nu}]q_{\nu}\right]u_s(k_2).
\label{qgf2}
\end{equation}
We  take
\begin{equation}
u_{s}(\vec{k})=
\sqrt{E_{\vec k}+m_f}\begin{pmatrix}
1 \\
\\
\dfrac{\vec{k}\cdot \vec{\sigma}}{E_{\vec k}+m_f}
\end{pmatrix} \chi_{s}
\label{qgf3}
\end{equation}
where
\begin{equation}
    \chi_+=\begin{pmatrix}
1 \\
0
\end{pmatrix}, \qquad \chi_-= \begin{pmatrix}
0 \\
1
\end{pmatrix}.
\label{qgf4}
\end{equation}
We next compute using Eqs.~\ref{qgf3}, \ref{qgf4}, in the non-relativistic limit,
\begin{equation}
\bar{u}_{s'}(\vec{k'}_2){u}_{s}(\vec{k}_2) =2m_f \left[\delta_{ss'} -\frac{i}{2m_f^2}(\vec{k}\times \vec{q})\cdot \vec{S}^{ss'}_{1/2}\right] + \ {\rm subleading~terms}
\label{qgf5}
\end{equation}
where we have defined the matrix elements of the spin vector as  ($\hbar=1$),
$$
\vec{S}^{ss'}_{1/2}=\frac12 \chi^{\dagger}_{s'} \vec{\sigma}\chi_{s}.
$$

\noindent
The three graviton vertex reads, e.g.~\cite{Holstein:2008sx} and also references therein,
\begin{eqnarray}
    \begin{split}
    \label{qg25}
        V^{(3)\mu\nu}_{\alpha\beta \gamma\delta}(k, q) =& -\frac{i\kappa}{2} \Big[  \mathcal{P} _{\alpha\beta \gamma\delta}\Big(k^{\mu}k^{\nu}+(k-q)^{\mu}(k-q)^{\nu}+q^{\mu}q^{\nu} -\frac32 \eta^{\mu\nu}q^2\Big) +2 q_{\lambda} q_{\sigma} \Big(I_{\alpha\beta}{}^{\sigma\lambda}I_{\gamma\delta}{}^{\mu\nu}+I_{\gamma\delta}{}^{\sigma\lambda}I_{\alpha\beta}{}^{\mu\nu} \\
        & -I_{\alpha\beta}{}^{\mu\sigma}I_{\gamma\delta}{}^{\nu\lambda}- I_{\gamma\delta}{}^{\mu\sigma}I_{\alpha\beta}{}^{\nu\lambda}  \Big) +\Big\{q_{\lambda} q^{\mu}\Big(\eta_{\alpha\beta} I_{\gamma\delta}{}^{\nu\lambda}+\eta_{\gamma\delta} I_{\alpha\beta}{}^{\nu\lambda} \Big)   + q_{\lambda} q^{\nu}\Big(\eta_{\alpha\beta} I_{\gamma\delta}{}^{\mu\lambda}+\eta_{\gamma\delta} I_{\alpha\beta}{}^{\mu\lambda} \Big) \\
        & - q^2 \Big(\eta_{\alpha\beta} I_{\gamma\delta}{}^{\mu\nu}+\eta_{\gamma\delta} I_{\alpha\beta}{}^{\mu\nu}\Big) - \eta^{\mu\nu}q_{\sigma}q_{\lambda}\Big(\eta_{\alpha \beta}I_{\gamma\delta}{}^{\sigma\lambda}+ \eta_{\gamma\delta}I_{\alpha \beta}{}^{\sigma\lambda}\Big)  \Big\} + \Big\{-2q_{\lambda} \Big(I_{\alpha\beta}{}^{\lambda \sigma} I_{\gamma\delta \sigma}{}^{\nu} (k-q)^{\mu} \\
        & +I_{\alpha\beta}{}^{\lambda \sigma} I_{\gamma\delta \sigma}{}^{\mu} (k-q)^{\nu} +  I_{\gamma\delta}{}^{\lambda \sigma}I_{\alpha\beta\sigma}{}^{\nu}k^{\mu}+ I_{\gamma\delta}{}^{\lambda \sigma}I_{\alpha\beta\sigma}{}^{\mu}k^{\nu} \Big) +q^2 \Big(I_{\alpha\beta\sigma}{}^{\mu}I_{\gamma\delta}{}^{\nu\sigma}+ I_{\gamma\delta\sigma}{}^{\mu}I_{\alpha\beta}{}^{\nu\sigma} \Big) \\
        & +\eta^{\mu\nu}q_{\sigma}q_{\lambda}\Big(I_{\alpha\beta}{}^{\lambda\rho}I_{\gamma\delta \rho}{}^{\sigma}+ I_{\gamma\delta}{}^{\lambda\rho}I_{\alpha\beta\rho}{}^{\sigma} \Big)  \Big\} + \Big\{\Big(k^2+(k-q)^2 \Big) \Big(I_{\alpha\beta}{}^{\mu\sigma}I_{\gamma\delta\sigma}{}^{\nu} + I_{\gamma\delta}{}^{\mu\sigma}I_{\alpha\beta\sigma}{}^{\nu} \\
        & -\frac12 \eta^{\mu\nu}{\cal P}_{\alpha\beta\gamma\delta}\Big)- \Big(I_{\gamma\delta}{}^{\mu\nu}\eta_{\alpha\beta}k^2+I_{\alpha\beta}{}^{\mu\nu}\eta_{\gamma\delta}(k-q)^2  \Big)  \Big\} \Big]
    \end{split}
\end{eqnarray}
where $\mathcal{P} _{\alpha\beta \gamma\delta}$ is given by Eq.~\ref{nmadd4}.
\begin{figure}[h!]
\begin{center}
    \includegraphics[width=0.2\linewidth]{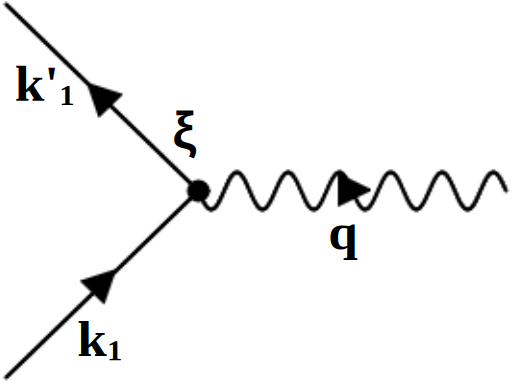} 
    \qquad  \includegraphics[width=0.15\linewidth]{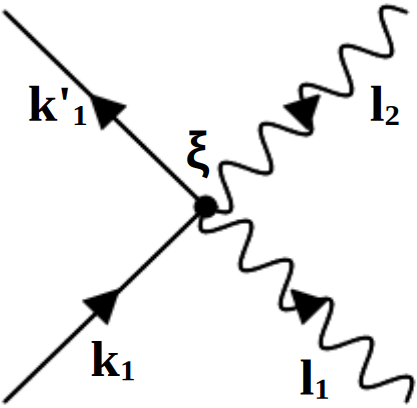} 
  \caption{\small \it The one and two graviton-two scalar non-minimal vertices. The dark circle on the junction represents that these vertices are non-minimal, compared to the minimal ones.   }
  \label{fnm}
\end{center}
\end{figure}

\noindent
Finally, we come to the issue of the non-minimal vertices { ($\xi \neq 0$),  \ref{fnm}. For one graviton-two scalar interaction, the relevant part of the  action reads,
\begin{eqnarray}
    S^{(1)}_{\xi}  = \frac{\xi \kappa}{2}\int d^4 x \ \eta_{\mu \nu}  \phi^2\partial^2 h^{\mu \nu}
    \label{nm1}
\end{eqnarray}
The corresponding vertex function reads
\begin{eqnarray}
        V_{({\xi})}^{\alpha \beta} (q) 
        = -i \xi \kappa  \eta^{\alpha \beta} q^2.
        \label{nmv1}
\end{eqnarray}
Note that the momentum appearing above is carried by the graviton, owing to the derivative acting on the same, Eq.~\ref{nm1}. As we mentioned earlier, this is opposite to the minimal vertex, as the latter contains derivatives of only the scalar field and not of the graviton. The part of the action relevant  for the two scalar-two graviton interaction reads 
\begin{eqnarray}
        S^{(2)}_{\xi} = \frac{\xi \kappa^2 }{2}\int d^4 x \ \phi^2 \Bigg[ \frac14   h \partial ^2 h + \frac12    (\partial ^{\lambda}h^{\mu\alpha})(\partial _{\alpha}h_{\mu\lambda})  -    h^{\mu\nu} \partial ^2 h_{\mu\nu} -\frac34   (\partial ^{\lambda} h^{\mu\nu})(\partial _{\lambda}h_{\mu\nu})\Bigg]  
\label{nmv2}
\end{eqnarray}
The two graviton-two scalar non-minimal vertex function reads,
\begin{eqnarray}
    \begin{split}
        V_{(\xi)}^{\mu \nu; \rho \sigma} (l_1,l_2) = - \frac{i \xi \kappa^2 }{4}  \Big[ (l_1^2 +l_2^2) (\eta^{\mu \nu} \eta^{\rho \sigma} - 4 \eta^{\mu \rho} \eta^{\nu \sigma}) - 2 ( l_1^{\sigma} l_2^{\nu } + l_1^{\nu} l_2^{\sigma } ) \eta^{\mu \rho} + 6 l_1\cdot l_2   \eta^{\mu \rho } \eta^{ \nu \sigma }   \Big].
    \end{split}
\end{eqnarray}

\noindent
As we have already mentioned earlier in \ref{S1},  there is a strong physical motivation to include a non-minimal interaction for a scalar owing to the issue of renormalisability~\cite{Parker:2009uva}. However, to the best of our knowledge, there is no sufficient purpose to consider such interactions for vector bosons or fermions. Hence, in order to maintain the simplest possible scenario, we will ignore any non-minimal interaction for non-spin-0 fields. In other words, $\xi$ will affect the scattering of scalar and non-spin-0 fields via the exchange of gravitons only.\\
With these  equipments, we are ready  for our scattering computations.

\section{The massive spin-0-massive spin-0 non minimal scattering and gravitational potential}\label{scalar}
The scattering between two massive non-minimal scalars in the presence of a three point vertex due to a cosmological constant (originating from the $-2\Lambda \sqrt{-g}/\kappa^2$ term in the action) in a scale much smaller compared to the Hubble horizon was computed recently in~\cite{Majumder:2025gou}. For $\Lambda=0$, it was also argued that the non-minimal coupling will result in a subleading long range gravitational potential compared to that of $\xi=0$, but no explicit results were presented. We wish to find out in this section  these explicit results for $\Lambda=0$.
 \begin{figure}[h]
         \begin{minipage}[b]{0.3\linewidth}
             \centering
             \includegraphics[width=0.8\linewidth]{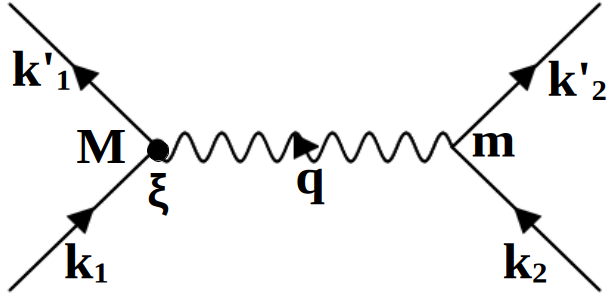}
          \end{minipage} 
         \begin{minipage}[b]{0.3\linewidth}
             \centering
             \includegraphics[width=0.6\linewidth]{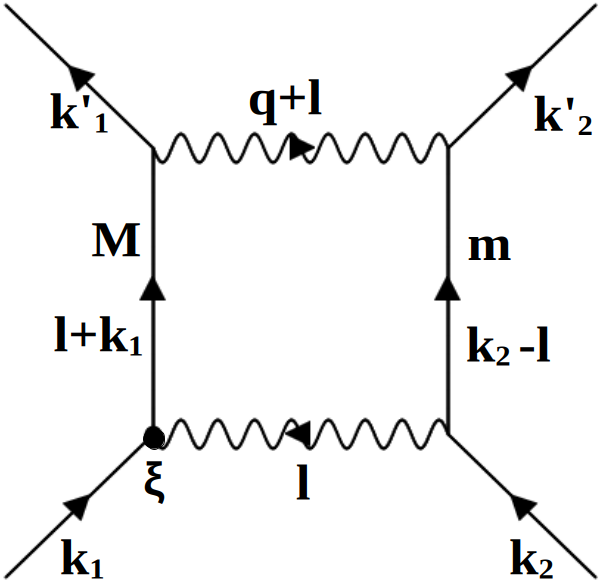}
         \end{minipage}
         \begin{minipage}[b]{0.3\linewidth}
             \centering
             \includegraphics[width=0.6\linewidth]{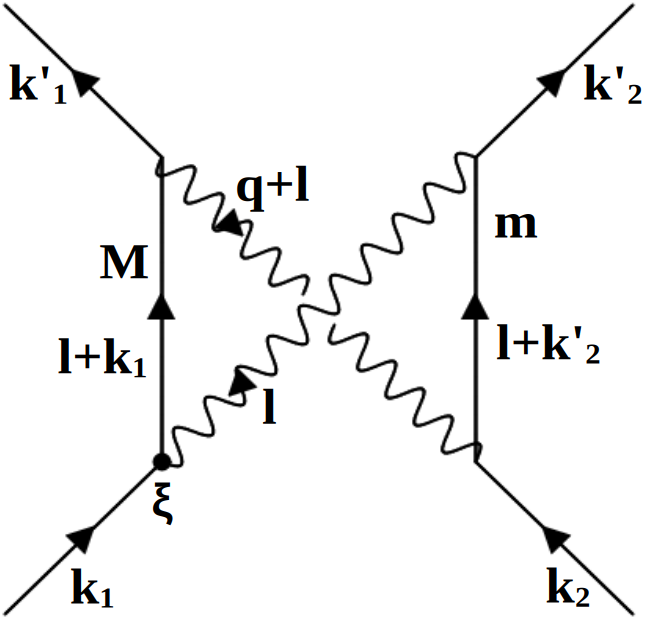}
          \end{minipage} 
           \caption{\it \small The tree, ladder and cross-ladder diagrams at linear order in the non-minimal coupling parameter $\xi$.  Hence the ladder and cross-ladder diagrams have four sub-categories each, depending on the placement of the $\xi$-vertex, denoted by the thick circle. The tree diagram has two sub-categories.}
           \label{tr-l-crl}
    \end{figure}

\subsection*{\small a) The tree diagrams :}

\noindent 
 Let us begin with the tree diagram, given by the first of Eqs.~\ref{tr-l-crl}. Using Eqs.~\ref{nmadd3}, \ref{qg5'}, \ref{nmv1}, the amplitude reads 
\begin{eqnarray}
&& i  \mathcal{M}^{\text{spin-0-spin-0}}_{\text{tree}}=-i (-i\kappa \xi) \left( - \frac{i\kappa}{2}\right) \frac{q^2 \eta^{\alpha\beta}{\cal P}_{\mu\nu\alpha\beta}\left(k_2^{\mu}{k'}^{\nu}_2+k_2^{\nu}{k'}^{\mu}_2-\eta^{\mu\nu}(k_2\cdot k'_2+m^2)\right) }{q^2}= i\kappa^2\xi m^2,
\label{t1}
\end{eqnarray}
where we have used the non-relativistic limit, $k_2\cdot k'_2 \approx -m^2$. 
Since this amplitude is independent of the transfer momentum $q^2$, its Fourier transform  is proportional to $\delta^3(\vec{r})$, and hence it does {\it not} contribute to any long range gravitational potential. Similar conclusion holds if we instead take the $k_2-k'_2$ vertex to be non-minimal.

\subsection*{\small b) The ladder and cross-ladder diagrams : }

\noindent
The ladder diagrams are given by the second of \ref{tr-l-crl},  having four sub-categories  depending upon whether the $\xi$-vertex 
is placed on the $k_1$ or $k'_1$ or $k_2$ or $k'_2$ lines. For the first, using $k_1\cdot k'_1\approx -M^2$, $k_2\cdot k'_2\approx -m^2$, we have 
\begin{eqnarray}
&& i  \mathcal{M}^{\text{spin-0-spin-0}}_{\text{ladder,1}}=\frac{\xi \kappa^4}{2^3}\eta_{\mu\nu}{\cal P}^{\mu\nu\alpha\beta}\left\{ k'_{2(\alpha} (k_2-l)_{\beta)} + \eta_{\alpha\beta} k'_2\cdot l \right\}   \nonumber\\&& \times
\frac{\left\{k'_{1(\lambda} (l+k_1)_{\rho)} - \eta_{\rho\lambda} k_1\cdot l  \right\} {\cal P}^{\lambda \rho \gamma \delta} \left\{  k_{2(\gamma} (k_2-l)_{\delta)} + \eta_{\gamma\delta} k_2\cdot l\right\} }
{(l+q)^2 [(l-k_2)^2+m^2] [(l+k_1)^2+M^2]}  \cdot 
\label{t2}
\end{eqnarray}
It is easy to see that the above amplitude also has no terms non-analytic in the transfer momentum $q^2$, and hence it has no contribution to long range gravitational potential. The same conclusion also holds for the three other ladder diagrams.\\

\noindent
Let us now come to the cross-ladder diagrams, the third of \ref{tr-l-crl}, which also has four sub-categories as of the ladder diagram. For the $\xi$-vertex placed upon the $k'_1$ line, we have 
\begin{eqnarray}
&& i  \mathcal{M}^{\text{spin-0-spin-0}}_{\text{cross-ladder,1}}=\frac{\xi \kappa^4}{2^3}\eta_{\mu\nu}{\cal P}^{\mu\nu\alpha\beta}\left\{ k'_{2(\alpha} (k'_2 + l)_{\beta)} + \eta_{\alpha\beta} k'_2\cdot l \right\}   \nonumber\\&& \times
\frac{\left\{k'_{1(\lambda} (l+k_1)_{\rho)} - \eta_{\rho\lambda} k'_1\cdot l  \right\} {\cal P}^{\lambda \rho \gamma \delta} \left\{  k_{2(\gamma} (k'_2 +l)_{\delta)} + \eta_{\gamma\delta} k_2\cdot l\right\} }
{(l+q)^2 [(l+ k'_2)^2+m^2] [(l+k_1)^2+M^2]},
\label{t3}
\end{eqnarray}
which also does not contribute to the long range potential. Likewise the other cross-ladder sub-categories do not contribute to the same. This was first argued in~\cite{Majumder:2025gou}.

\subsection*{\small c) The triangle diagrams :}
There are total six triangle diagrams given in \ref{f3}. Note that  for the first four of them the non-minimal vertices  are three point, whereas for the last two, they are four point. 
 \begin{figure}[h]
         \begin{minipage}[b]{0.3\linewidth}
             \centering
             \includegraphics[width=0.6\linewidth]{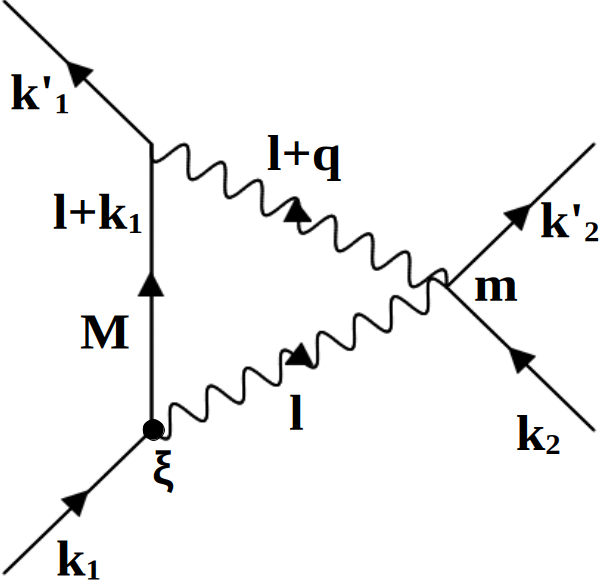}
          \end{minipage} 
         \begin{minipage}[b]{0.3\linewidth}
             \centering
             \includegraphics[width=0.6\linewidth]{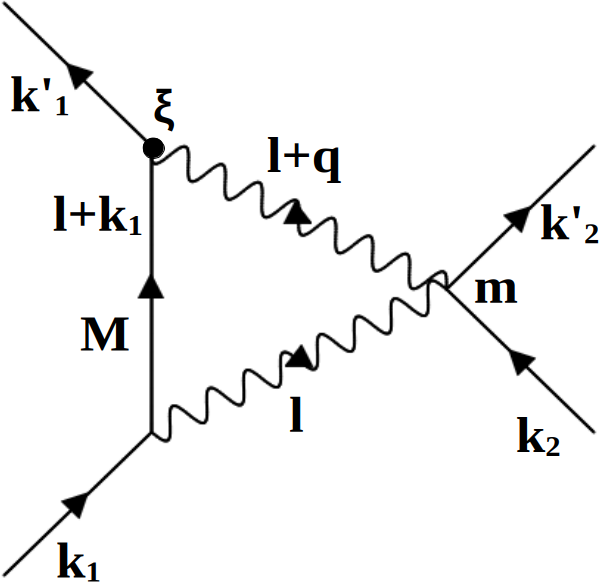}
         \end{minipage}
         \begin{minipage}[b]{0.3\linewidth}
             \centering
             \includegraphics[width=0.6\linewidth]{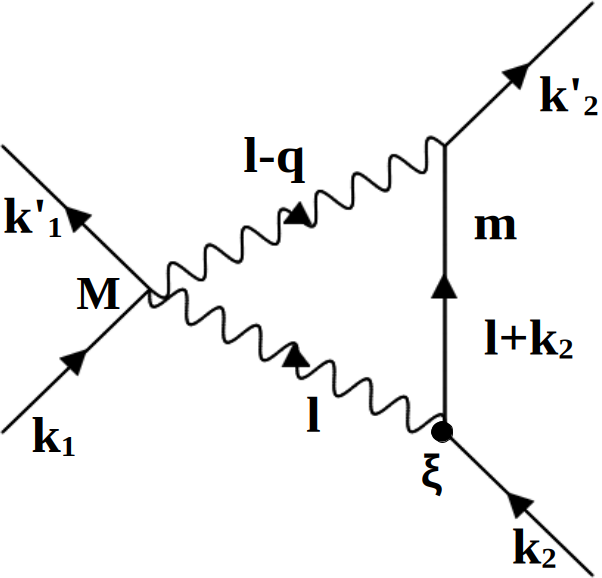}
          \end{minipage} \\
          \\
         \begin{minipage}[b]{0.3\linewidth}
             \centering
             \includegraphics[width=0.6\linewidth]{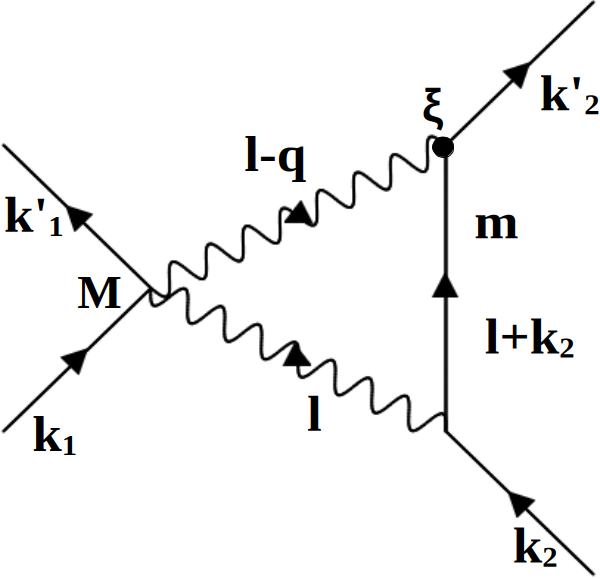}
         \end{minipage}
         \begin{minipage}[b]{0.3\linewidth}
             \centering
             \includegraphics[width=0.6\linewidth]{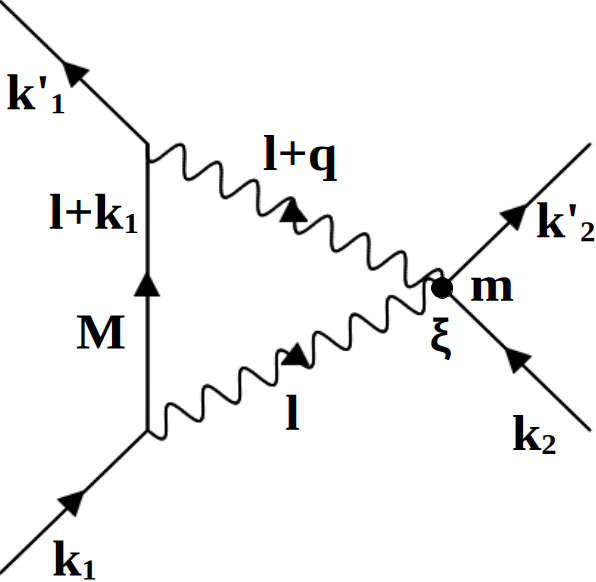}
          \end{minipage} 
         \begin{minipage}[b]{0.3\linewidth}
             \centering
             \includegraphics[width=0.6\linewidth]{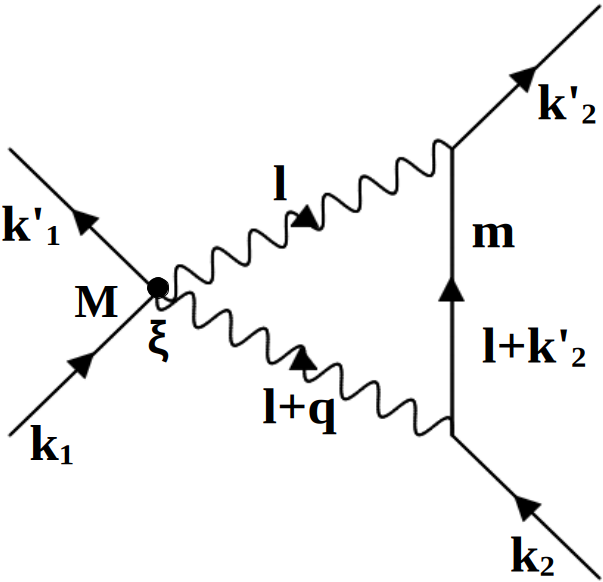}
         \end{minipage}
         \caption{\it \small The triangle diagrams for massive spin-0-spin-0 fields non-minimal  scattering at ${\cal O}(G^2\xi)$.}
         \label{f3}
 \end{figure}
The Feynman amplitude for the first diagram reads,
\begin{eqnarray}
    \begin{split}
        i \mathcal{M}^{\text{spin-0-spin-0}}_{\text{Triangle-1}} =& \int \frac{d^4 l}{(2 \pi)^4} V^ {\text{spin-0}\,(1)}_{\mu  \nu \,(\xi)} (l) V^{\text{spin-0\,(1)}}_{\alpha \beta} (l+k_1,k_1',M) \frac{{- i \cal P}^{\mu \nu \phi \lambda}}{l^2} \frac{{- i \cal P}^{\alpha \beta \rho \sigma}}{(l+q)^2} V^{\text{spin-0\,(2)}}_{\,\phi \lambda \rho \sigma} (k_2,k_2',m) \\
        & \times\frac{-i}{[(l+k_1)^2 + M^2]} \\
    \end{split}
\end{eqnarray}
It is easy to check that the  above reduces to an integral like
 $$\sim \int \dfrac{d^4 l}{(2 \pi)^4}\dfrac{1}{l^2 [(l+k_1)^2 + M^2]},$$
  which makes no  contributions  non-analytic in the transfer momentum squared, $q^2$. A similar conclusion holds for  the second, third, and fourth triangle diagrams.

The last two diagrams containing four point non-minimal vertices Eq.~\ref{nmv2} contribute to our present purpose as follows. Using the integrals given in \ref{A}, the Feynman amplitudes respectively read,
\begin{eqnarray}
    \begin{split}
        i \mathcal{M}^{\text{spin-0-spin-0}}_{\text{Triangle-5}} =& \int \frac{d^4 l}{(2 \pi)^4}  V^{\text{spin-0\,(1)}}_{\mu \nu} (k_1,l+k_1,M) V^{\text{spin-0\,(1)}}_{\alpha \beta} (l+k_1,k_1',M) \frac{{- i \cal P}^{\mu \nu \phi \lambda}}{l^2} \frac{{- i \cal P}^{\alpha \beta \rho \sigma}}{(l+q)^2}V^{\text{spin-0\,(2)}}_{\phi \lambda \rho \sigma \,(\xi)} (l+q,l) \\
        & \times \frac{-i}{[(l+k_1)^2 + M^2]} \\
 =& \,i\,G^2 \xi \bigg( \frac{2  q^6 \ln q^2}{M^2} +18  q^4 \ln q^2 -\frac{140}{3} M^2  q^2 \ln q^2 +\frac{\pi ^2  q^5}{M}+\frac{27}{2} \pi ^2 M  q^3  +\frac{16}{3} M^2  q^2 -32 \pi ^2 M^3  q \bigg),
    \end{split}
    \label{potamp1}
\end{eqnarray}
and, 
\begin{eqnarray}
    \begin{split}
        i \mathcal{M}^{\text{spin-0-spin-0}}_{\text{Triangle-6}} =& \int \frac{d^4 l}{(2 \pi)^4}  V^{\text{\:spin-0\,(2)}}_{\phi \lambda \rho \sigma \,(\xi)} (l,l+q) \frac{{- i \cal P}^{ \phi \lambda \mu \nu}}{l^2} \frac{{- i \cal P}^{ \rho \sigma \alpha \beta }}{(l+q)^2} V^{\text{\:spin-0\,(1)}}_{\mu \nu} (l+ k_2, k_2',m) V^{\text{spin-0\,(1)}}_{\alpha \beta} (k_2,l+k_2,m) \\
        & \times \frac{-i}{[(l+k_2')^2 + m^2]} \\
         =&\,i\, G^2 \xi \bigg(\frac{2  q^6 \ln q^2}{m^2} + 18  q^4 \ln q^2 -\frac{140}{3} m^2  q^2 \ln q^2 +\frac{ \pi ^2  q^5}{ m} + \frac{ 27 }{2} \pi ^2 m  q^3 +\frac{16}{3} m^2  q^2 - 32 \pi ^2 m^3  q \bigg),
    \end{split}
    \label{potamp2}
\end{eqnarray}
where $q^{2n+1}= (q^2)^{(2n+1)/2}$ is understood.
We need to retain only the  pieces that are non-analytic in  $q^2$. In the non-relativistic limit we take $q \approx \{ 0, \vec{q}\}$. The gravitational potential is then defined as the Fourier transform
$$V(\vec{r}) = -\frac{1}{4M  m}\int \frac{d^3 \vec{q}}{(2\pi)^3}e^{-i \vec{q}\cdot \vec{r}} \ \mathcal{M}(\vec{q}^{\,\,2})\vert_{\rm non-analytic}.$$

\noindent
Using then the list written in \ref{A}, the long range gravitational potentials corresponding to Eqs.~\ref{potamp1}, \ref{potamp2}, respectively reads
\begin{equation}
        \begin{split}
        &V^{\text{spin-0-spin-0}}_{\text{Triangle-5}} (G^2,r, \xi) = \frac{G^2 \xi}{ m r^4} \bigg(- 8 M^2  +\frac{35 M }{\pi  r}-\frac{81 }{2 r^2} +\frac{270 }{\pi M r^3} -\frac{90 }{ M^2 r^4}+ \frac{1260 }{\pi M^3 r^5} \bigg),
    \\
        &V^{\text{spin-0-spin-0}}_{\text{Triangle-6}}(G^2,r, \xi) = \frac{G^2 \xi}{ M r^4} \bigg( - 8 m^2  +\frac{35 m }{\pi r} - \frac{  81 }{2 r^2} + \frac{ 270 }{\pi  m  r^3} -\frac{90  }{m^2  r^4} + \frac{ 1260 }{\pi  m^3  r^5} \bigg).
    \end{split}
    \label{pot1}
\end{equation}
Note that the positions of $m$ and $M$ are flipped in the two above potentials, as one expects from the topology of the respective diagrams. Note also that the terms originating from the $\ln \vec{q}^{\,\,2}$ should be interpreted as quantum.


\subsection*{\small d) The seagull diagrams :}
\begin{figure}[ht]
        \begin{minipage}[b]{0.3\linewidth}
            \centering
            \includegraphics[width=0.6\linewidth]{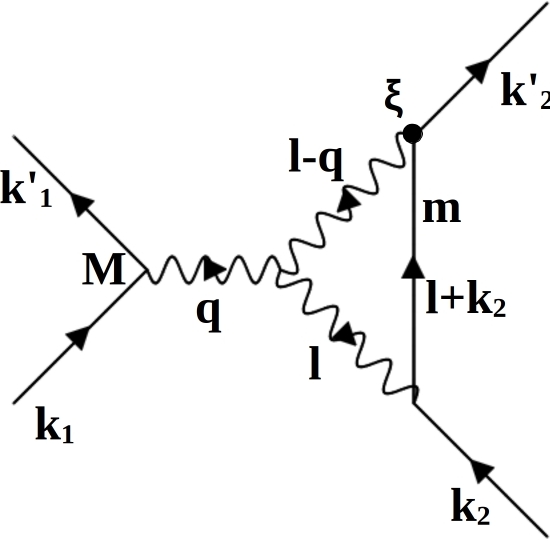}
         \end{minipage} 
        \begin{minipage}[b]{0.3\linewidth}
            \centering
            \includegraphics[width=0.6\linewidth]{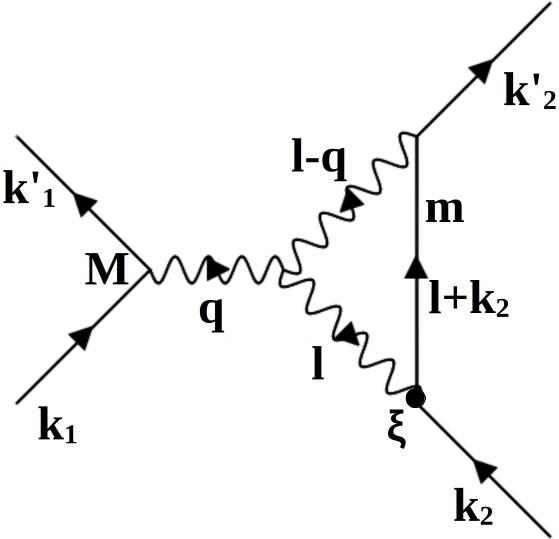}
        \end{minipage}
        \begin{minipage}[b]{0.3\linewidth}
            \centering
            \includegraphics[width=0.6\linewidth]{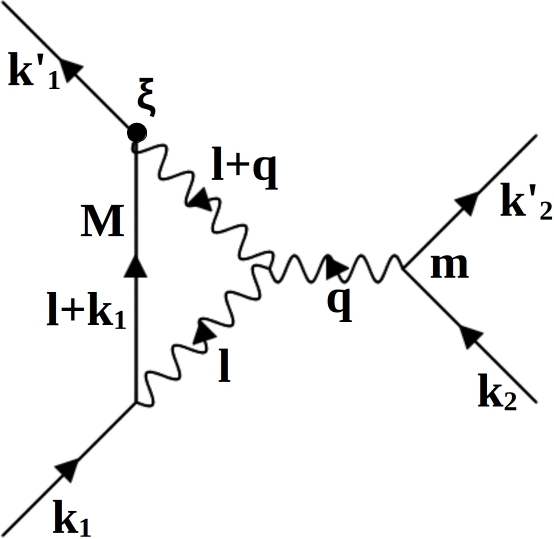}
         \end{minipage}\\
         \\
        \begin{minipage}[b]{0.3\linewidth}
            \centering
            \includegraphics[width=0.6\linewidth]{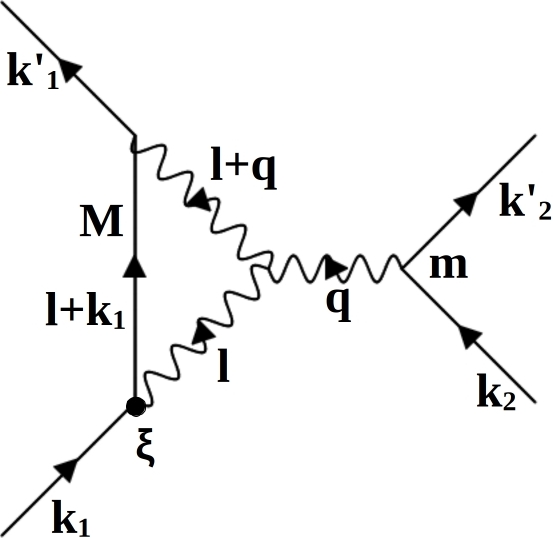}
        \end{minipage}
        \begin{minipage}[b]{0.3\linewidth}
            \centering
            \includegraphics[width=0.6\linewidth]{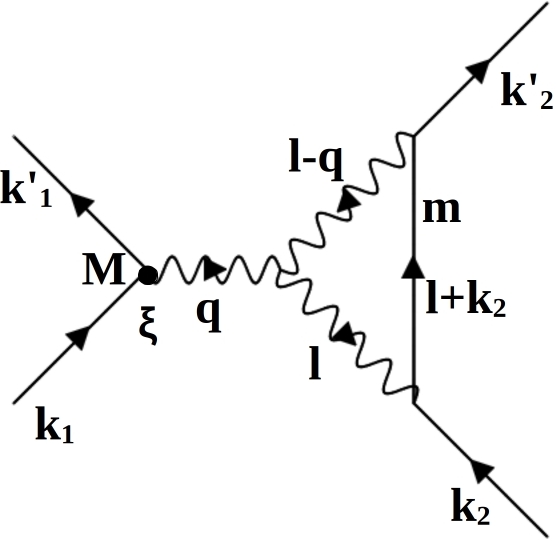}
         \end{minipage} 
        \begin{minipage}[b]{0.3\linewidth}
            \centering
            \includegraphics[width=0.6\linewidth]{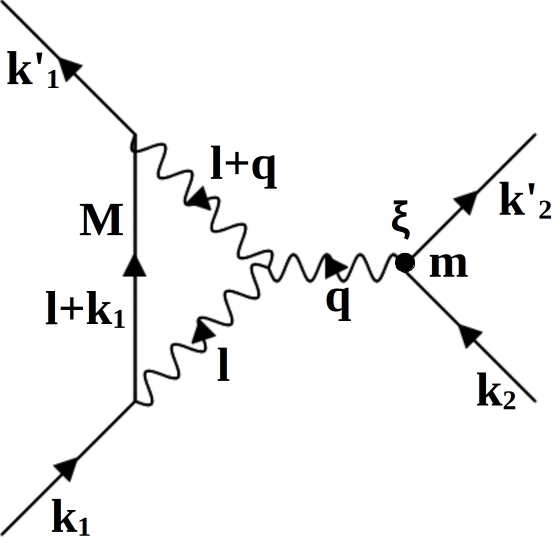}
        \end{minipage}
        \caption{\it \small The seagull diagrams for massive spin-0-spin-0 fields scattering at ${\cal O}(G^2\xi)$.}
        \label{f4}
\end{figure}
There are total six seagull diagrams as depicted in \ref{f4}. Note the two  groups of different topologies  here,  i.e., diagrams one to four and five to six.  The Feynman amplitude for the first diagram is, 
\begin{eqnarray}
    \begin{split}
        i \mathcal{M}^{\text{spin-0-spin-0}}_{\text{seagull-1}} =& \int \frac{d^4 l}{(2 \pi)^4} V_{\text{spin-0} \,(1)}^{\mu \nu} (k_1,k_1',M) \frac{-i \cal P_{\mu \nu \alpha \beta}}{q^2} V^{ \alpha \beta \, (3)} _{ \gamma \delta \rho \sigma} (l-q,-q) \frac{-i \cal P^{\gamma \delta \lambda \phi}}{(l-q)^2}  \frac{-i \cal P^{\rho \sigma \psi \theta}}{l^2} V^ {\text{spin-0}\,(1)}_{\lambda \phi \,(\xi)} (l-q)  \\
        &\times V^{\text{spin-0}\,(1)}_{\psi \theta} (k_2,l + k_2,m) \frac{- i}{[(l+k_2)^2 +m^2]}, 
        \label{sg1}
    \end{split}
\end{eqnarray}
which reduces to an integral like 
$$\sim \frac{1}{q^2} \int \dfrac{d^4 l}{(2 \pi)^4}\dfrac{1}{l^2 [(l+k_2) + m^2]}=\frac{i}{16 \pi^2 q^2} \bigg[ \frac{2}{\epsilon} - \ln \frac{m^2}{4 \pi \mu^2} +2 (\ln2-1)  \bigg]. $$
Note that the $q^2$ appearing above comes from the graviton propagator, and it gets factorised with the rest of the amplitude, which is basically the 1PI one loop correction (${\cal O}(\kappa^3 \xi)$)  of the three point non-minimal  vertex. The above contribution coming from this vertex function is just a constant, and hence it can be absorbed in a vertex counterterm. Thus Eq.~\ref{sg1} does not contribute to the gravitational potential. A similar conclusion holds for the seagull diagrams 2, 3 and 4. The situation is, however, different for the fifth and sixth seagull diagrams. The corresponding amplitudes respectively read,
\begin{eqnarray}
    \begin{split}
        i \mathcal{M}^{\text{spin-0-spin-0}}_{\text{seagull-5}} =& \int \frac{d^4 l}{(2 \pi)^4} V_{\text{spin-0} \,(1)}^{\rho \sigma \,(\xi)} (q) \frac{-i \cal P_{\rho \sigma \mu \nu }}{q^2} V^{\mu \nu \,(3)  } _{ \alpha \beta \gamma \delta } (l-q,-q) \frac{-i \cal P^{\gamma \delta \psi \theta }}{l^2}  \frac{-i \cal P^{\alpha \beta \lambda \phi}}{(l-q)^2}  V^{\text{spin-0}\,(1)}_{ \psi \theta} (k_2,l + k_2,m)\\
        & \times V^ {\text{spin-0}\,(1)}_{ \lambda \phi} (l+ k_2,k_2',m)  \frac{- i}{[(l+k_2)^2 +m^2]} \\
        =& \,i\, G^2 \xi \,\Bigg[ q^2 \log q^2\Bigg( \frac{16 q^4}{m^2} +48 \, q^2 + 72  m^2  \Bigg) + \frac{6 \pi ^2  q^5}{m} + 14 \pi ^2  m \,  q^3   +   48 \pi ^2  m^3 \,  q \Bigg],
    \end{split}
\end{eqnarray}
and,
\begin{eqnarray}
    \begin{split}
        i \mathcal{M}^{\text{spin-0-spin-0}}_{\text{seagull-6}}=& \int \frac{d^4 l}{(2 \pi)^4} V^ {\text{spin-0}\,(1)}_{ \lambda \phi} (l+ k_1,k_1',M)  V^{\text{spin-0}\,(1)}_{ \psi \theta} (k_1,l + k_1,M) \frac{-i \cal P^{\psi \theta \gamma \delta }}{l^2}  \frac{-i \cal P^{\lambda \phi \alpha \beta }}{(l+q)^2} V^{\mu \nu \,(3)  } _{ \alpha \beta \gamma \delta } (l+q,q) \\
        & \times \frac{-i \cal P_{\mu \nu \rho \sigma}}{q^2}   V_{\text{spin-0} \,(1)}^{\rho \sigma \,(\xi)} (q) \frac{- i}{[(l+k_1)^2 +M^2]} \\
=& \,i\,G^2 \xi \bigg[ q^2\, \ln q^2 \bigg( \frac{16   q^4 }{M^2} +48  q^2  +72  M^2 \bigg) +\frac{6 \pi ^2   q^5}{M}+14 \pi ^2  M  \, q^3  + 48 \pi ^2  M^3 \,q\bigg].
    \end{split}
\end{eqnarray}
The corresponding potentials are,
\begin{equation}
    \begin{split}
        V^{\text{spin-0-spin-0}}_{\text{seagull-5}} (G^2,r, \xi) =&  \frac{6 G^2 \xi}{M r^4} \bigg(  2 m^2  - \frac{9 m }{\pi r}-\frac{7 }{ r^2} +\frac{ 120 }{\pi  m r^3} - \frac{ 90 }{m^2 r^4}  + \frac{ 1680 }{\pi  m^3 r^5} \bigg), \\
        V^{\text{spin-0-spin-0}}_{\text{seagull-6}} (G^2,r, \xi)=& \frac{6G^2 \xi}{m r^4} \bigg( 2 M^2  -\frac{9 M }{\pi r}-\frac{7 }{ r^2} +\frac{120 }{\pi M r^3} -\frac{90 }{ M^2 r^4} + \frac{ 1680 }{\pi M^3 r^5} \bigg).
    \end{split}
    \label{sg}
\end{equation}
Note once again as a consistency check the symmetry under the interchange of $m$ and $M$ in the above potentials. 

\subsection*{\small  e) The double seagull diagrams :}
Let us now come to the two double seagull diagrams given by \ref{f5}.
\begin{figure}[ht]
        \hspace{1cm}\begin{minipage}[b]{0.3\linewidth}
            \centering
            \includegraphics[width=0.8\linewidth]{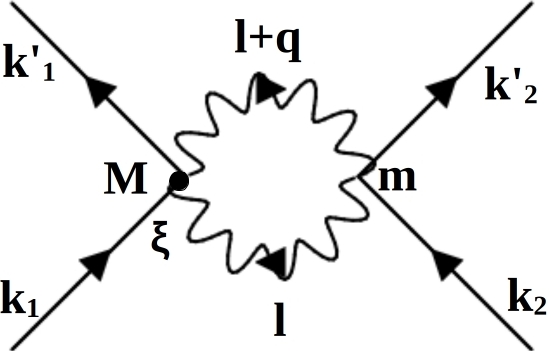}
         \end{minipage} 
        \hspace{4cm}
        \begin{minipage}[b]{0.3\linewidth}
            \centering
            \includegraphics[width=0.8\linewidth]{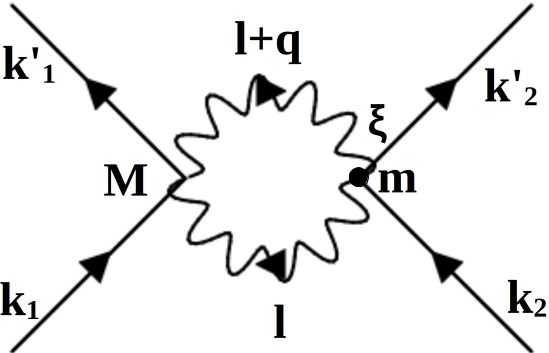}
        \end{minipage}
        \caption{\it \small The double seagull diagrams for massive spin-0-spin-0 fields scattering. }
        \label{f5}
\end{figure} 
Their Feynman amplitudes read, 
\begin{eqnarray}
    \begin{split}
             i \mathcal{M}^{\text{spin-0-spin-0}}_{\text{double seagull-1}} =&  \frac{1}{2!} \int \frac{d^4 l}{(2 \pi)^4} V_{\eta \lambda \rho \sigma \,(\xi)}^{\text{spin-0}\,(2)} (l,l+q) \frac{-i \mathcal{P}^{ \rho \sigma \mu \nu }}{(l+q)^2} \frac{-i \mathcal{P}^{\eta \lambda \alpha \beta}}{l^2} V^{\text{spin-0}\,(2)}_{ \alpha \beta \mu \nu } (k_2,k_2 ',m)  \\ 
         =& \,i\,\frac{40}{3} G^2 \xi  q^4 \ln q^2 + \,i\, \frac{584}{3} G^2 m^2 \xi  q^2 \ln q^2,
    \end{split}
\end{eqnarray}
and,
\begin{eqnarray}
    \begin{split}
            i \mathcal{M}^{\text{spin-0-spin-0}}_{\text{double seagull-2}} =& \frac{1}{2!} \int \frac{d^4 l}{(2 \pi)^4}  V^{\text{spin-0}\,(2)}_{ \eta \lambda \rho \sigma } (k_1,k_1 ',M)  \frac{-i \mathcal{P}^{\eta \lambda  \alpha \beta }}{l^2}  \frac{-i \mathcal{P}^{ \rho \sigma \mu \nu }}{(l+q)^2}    V_{\alpha \beta \mu \nu \,(\xi)}^{\text{spin-0}\,(2)}  (l+q,l)  \\ 
            =& \,i\,\frac{40}{3} G^2 \xi  q^4 \ln q^2 + \,i\,\frac{584}{3} G^2 M^2 \xi  q^2 \ln q^2.
    \end{split}
\end{eqnarray}
Accordingly, the corresponding contributions to the gravitational potential are evaluated to be,
\begin{equation}
    \begin{split}
        V^{\text{spin-0 - spin-0}}_{\text{double seagull-1}} (G^2,r, \xi) = \frac{2 G^2 \xi}{ \pi  M r^5} \bigg(- 73 m + \frac{100 }{ m r^2} \bigg), \hspace{0.5cm}
        V^{\text{spin-0 - spin-0}}_{\text{double seagull-2}}(G^2,r, \xi) = \frac{2 G^2 \xi}{ \pi  m r^5} \bigg(- 73 M + \frac{100 }{ M r^2} \bigg).
        \label{ds}
    \end{split}
\end{equation}
\subsection*{\small f) The fish diagrams :}
There are total four fish diagrams, \ref{f6}, all of which contribute to the long range gravitational potential. The corresponding Feynman amplitudes  read,
\begin{figure}[ht]
        \hspace{2cm}\begin{minipage}[b]{0.3\linewidth}
            \centering
            \includegraphics[width=0.8\linewidth]{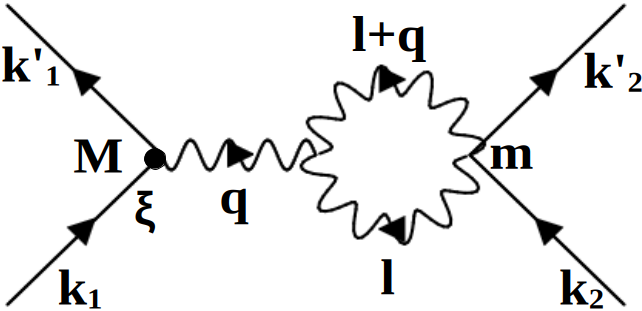}
         \end{minipage} 
        \hspace{3cm}
        \begin{minipage}[b]{0.3\linewidth}
            \centering
            \includegraphics[width=0.8\linewidth]{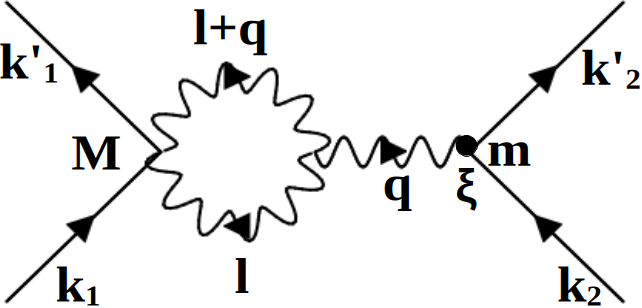}
        \end{minipage} \\
        \\
        \\
        \hspace*{2cm}\begin{minipage}[b]{0.3\linewidth}
            \centering
            \includegraphics[width=0.8\linewidth]{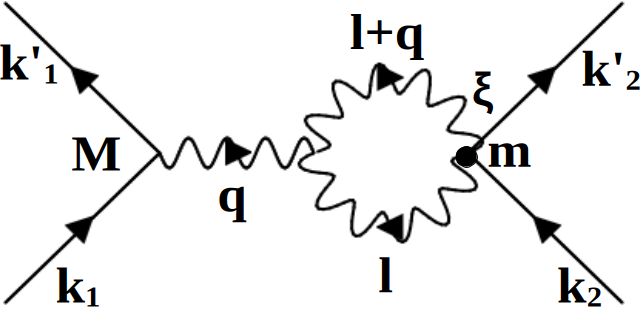}
         \end{minipage} 
        \hspace{3cm}
        \begin{minipage}[b]{0.3\linewidth}
            \centering
            \includegraphics[width=0.8\linewidth]{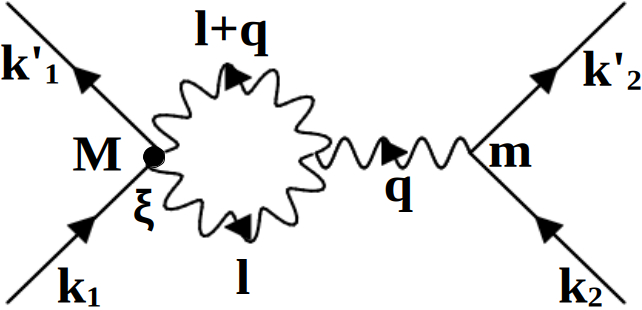}
        \end{minipage}
        \caption{\it \small The fish diagrams for massive spin-0-spin-0 fields scattering.}  
         \label{f6}
\end{figure} 
\begin{eqnarray}
    \begin{split}
         i \mathcal{M}^{\text{spin-0-spin-0}}_{\text{fish-1}} =& \frac{1}{2!} \int \frac{d^4 l}{(2 \pi)^4} V_{\text{spin-0}\,(1)}^{\lambda \phi\,(\xi)} (q) \frac{-i \mathcal{P}_{\lambda \phi \mu \nu}}{q^2} V^{\mu \nu\,(3)} _{\alpha \beta \gamma \delta} (l,-q) \frac{-i \mathcal{P}^{\alpha \beta \psi \theta}}{l^2}  \frac{-i \mathcal{P}^{\gamma \delta \rho \sigma}}{(l+q)^2} V^{\text{spin-0}\,(2)}_{\psi \theta \rho \sigma } (k_2,k_2',m)  \\
        =& -\,i 64 G^2 \xi  q^4 \ln q^2 -\,i 1136 G^2 m^2 \xi  q^2 \ln q^2,
    \end{split}
\end{eqnarray}
\begin{eqnarray}
    \begin{split}
          i \mathcal{M}^{\text{spin-0-spin-0}}_{\text{fish-2}} =& \frac{1}{2!} \int \frac{d^4 l}{(2 \pi)^4} V_{\rho \sigma \psi \theta} (k_1,k_1',M) \frac{-i \mathcal{P}^{\rho \sigma \gamma \delta }}{l^2} \frac{-i \mathcal{P}^{\psi \theta \alpha \beta }}{(l+q)^2} V^{\mu \nu\,(3)} _{\alpha \beta \gamma \delta} (l+q,q) \frac{-i \mathcal{P}_{\mu \nu \lambda \phi }}{q^2}  V_{\text{spin-0}\,(1)}^{\lambda \phi\,(\xi)} (q)  \\
        =& -\,i 64 G^2 \xi  q^4 \ln q^2 -\,i 1136 G^2 M^2 \xi  q^2 \ln q^2,
    \end{split}
\end{eqnarray}
\begin{eqnarray}
    \begin{split}
         i \mathcal{M}^{\text{spin-0-spin-0}}_{\text{fish-3}} =& \frac{1}{2!} \int \frac{d^4 l}{(2 \pi)^4} V_{\text{spin -0 (1)}}^{\lambda \phi} (k_1,k_1',m) \frac{-i \mathcal{P}_{\lambda \phi \mu \nu}}{q^2} V^{\mu \nu\,(3)} _{\alpha \beta \gamma \delta} (l,-q) \frac{-i \mathcal{P}^{\alpha \beta \psi \theta}}{l^2} \frac{-i \mathcal{P}^{\gamma \delta \rho \sigma}}{(l+q)^2}  V^{\text{spin-0\,(2)}}_{\rho \sigma \psi \theta\,(\xi)} (l+q,l)  \\
        =& -\, \frac{310i}{3}  G^2 \xi  q^4 \ln q^2 -\, \frac{2060 i}{3} G^2 M^2 \xi  q^2 \ln q^2,
    \end{split}
\end{eqnarray}
and,
\begin{eqnarray}
    \begin{split}
          i \mathcal{M}^{\text{spin-0-spin-0}}_{\text{fish-4}} =& \frac{1}{2!} \int \frac{d^4 l}{(2 \pi)^4} V^{\text{spin-0\,(2)}}_{\rho \sigma \psi \theta\,(\xi)} (l,l+q) \frac{-i \mathcal{P}^{ \rho \sigma \gamma \delta}}{l^2} \frac{-i \mathcal{P}^{\psi \theta \alpha \beta }}{(l+q)^2}  V^{\mu \nu\,(3)} _{\alpha \beta \gamma \delta} (l+q,q)  \frac{-i \mathcal{P}_{\mu \nu \lambda \phi }}{q^2} V_{\text{spin -0 (1)}}^{\lambda \phi} (k_2,k_2',m)  \\ 
            =& -\,\frac{310 i}{3}  G^2 \xi  q^4 \ln q^2 -\,\frac{2060i }{3} G^2 m^2 \xi  q^2 \ln q^2.
    \end{split}
\end{eqnarray}
Their respective contributions to the gravitational potential are given by,
\begin{equation}
    \begin{split}
        V^{\text{spin-0 - spin-0}}_{\text{fish-1}} (G^2,r, \xi) =& \frac{4 G^2 \xi}{ \pi M r^5} \bigg( 213 m -\frac{240 }{  m r^2} \bigg), \qquad 
        V^{\text{spin-0 - spin-0}}_{\text{fish-2}} (G^2,r, \xi) = \frac{4 G^2 \xi}{ \pi m r^5} \bigg( 213 M -\frac{240 }{  M r^2} \bigg), \\
        V^{\text{spin-0 - spin-0}}_{\text{fish-3}} (G^2,r, \xi)  =& \frac{5 G^2 \xi}{\pi m r^5} \bigg( 103  M-\frac{310  }{ M r^2} \bigg),\qquad 
        V^{\text{spin-0 - spin-0}}_{\text{fish-4}} (G^2,r, \xi) = \frac{5 G^2 \xi}{\pi M r^5} \bigg( 103 m-\frac{310 }{ m r^2} \bigg).
        \label{fish}
    \end{split}
\end{equation}
\subsection*{ \small g) The vacuum polarisation diagrams :}

Finally, we come to the two vacuum polarisation diagrams at ${\cal O}(G^2 \xi)$ as shown in \ref{f6}.
\begin{figure}[h]
        \hspace{2cm}\begin{minipage}[b]{0.3\linewidth}
            \centering
            \includegraphics[width=0.85\linewidth]{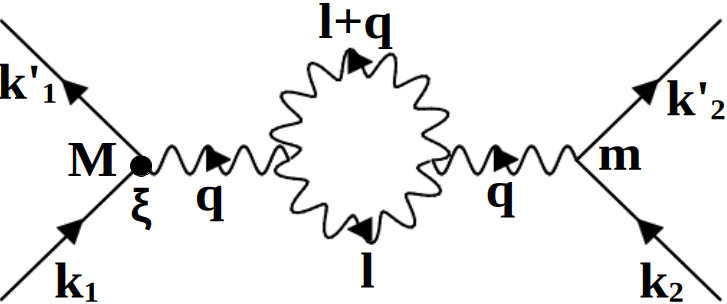}
         \end{minipage} 
        \hspace{3.5cm}
        \begin{minipage}[b]{0.3\linewidth}
            \centering
            \includegraphics[width=0.85\linewidth]{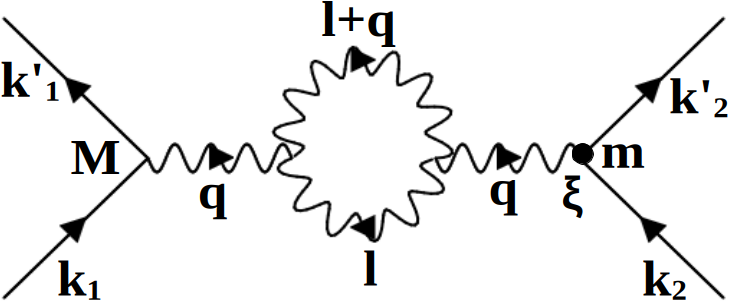}
        \end{minipage}
        \caption{\it \small The vacuum polarization diagrams for massive spin-0-spin-0 fields scattering. The contributions from the ghost loop needs also to be added. }
        \label{f6}
\end{figure}
The Feynman amplitudes are given by, 
\begin{eqnarray}
    \begin{split}
           \mathcal{M}^{\text{spin-0-spin-0}}_{\text{vac. pol.-1}} =& V^{\text{spin-0}\,(1)}_{\mu \nu\,(\xi)} (q)  \dfrac{ - i \mathcal{P}^{\mu \nu \rho \sigma }}{q^2} \Pi_{\rho \sigma \lambda \phi } (q) \ \dfrac{- i \mathcal{P}^{\lambda \phi \gamma \delta}}{q^2}   V^{\text{spin -0 (1)}}_{\gamma \delta} (k_2,k_2',m)  \\ 
            =& 12 G^2 \xi  q^4 \ln q^2 + 24 G^2 m^2 \xi  q^2 \ln q^2,
    \end{split}
\end{eqnarray}  
and,
\begin{eqnarray}
    \begin{split}
          \mathcal{M}^{\text{spin-0-spin-0}}_{\text{vac. pol. -2}} =& V^{\text{spin-0 (1)}}_{\mu \nu} (k_1,k_1',M)  \dfrac{ - i \mathcal{P}^{\mu \nu \rho \sigma }}{q^2} \Pi_{\rho \sigma \lambda \phi } (q) \ \dfrac{- i \mathcal{P}^{\lambda \phi \gamma \delta}}{q^2}  V^{\text{spin-0}\,(1)}_{\gamma \delta \,(\xi)} (q)  \\ 
            =& 12 G^2 \xi  q^4 \ln q^2 +24 G^2 M^2 \xi  q^2 \ln q^2,
    \end{split}
\end{eqnarray}  
where $ \Pi_{\alpha \beta \gamma \delta }$ is the one loop graviton self energy due to itself  after adding the ghost contribution, reading~\cite{tHooft:1974toh},
\begin{eqnarray}
\begin{split}
            \Pi_{\alpha \beta \gamma \delta } =& -\frac{2G}{\pi}\ln q^2 \Big[\frac{21}{120}q^4 I_{\alpha \beta \gamma \delta} + \frac{23}{120}q^4 \eta_{\alpha \beta}\eta_{\gamma \delta}- \frac{23}{120}q^2 (\eta_{\alpha \beta}q_{\gamma} q_{\delta} + \eta_{\gamma \delta} q_{\alpha} q_{\beta}) - \frac{21}{240}q^2 (q_{\alpha} q_ {\delta}\eta_{\beta \gamma} + q_{\beta} q_ {\delta}\eta_{\alpha \gamma} \\
            & + q_{\alpha} q_ {\gamma}\eta_{\beta \delta} + q_{\beta} q_ {\gamma}\eta_{\alpha \delta}) + \frac{11}{30}q_{\alpha} q_{\beta} q_{\gamma}q_{\delta}\Big].
        \end{split}
        \label{vpol}
\end{eqnarray}
The potentials for these two diagrams read 
\begin{equation}
    \begin{split}
        V^{\text{spin-0-spin-0}}_{\text{vac. pol.-1}} (G^2,r, \xi) =& \frac{18 G^2 \xi}{ \pi M r^5}\bigg(- m  + \frac{10 }{  m r^2} \bigg), \qquad 
        V^{\text{spin-0-spin-0}}_{\text{vac. pol.-2}} (G^2,r, \xi) = \frac{18 G^2 \xi}{ \pi m r^5}\bigg(- M  + \frac{10 }{  M r^2} \bigg).
        \label{vp}
    \end{split}
\end{equation}
\subsection{\small The full long range gravitational potential at ${\cal O}(G^2 \xi)$ :}
Combining now the different contributions from Eqs.~\ref{pot1}, \ref{sg}, \ref{ds}, \ref{fish}, \ref{vp}, we finally obtain the total long range gravitational potential at ${\cal O}(G^2 \xi)$,
\begin{eqnarray}
    \begin{split}
          V^{\text{spin-0-spin-0}} (G^2,r,\xi)
        =& \frac{4G^2 \xi}{r^4} \left[\left(\frac{M^2 }{m } +\frac{m^2 }{M }\right) +\left(\frac{M }{m} +\frac{  m }{M}\right) \frac{296}{\pi r} -\left(\frac{1}{m}+\frac{1}{M}\right)  \frac{165}{8r^2}\right. \\ 
        &\left.   -\frac{570}{\pi  m M r^3} 
       -\left( \frac{1}{m} +\frac{1}{M}\right)\frac{315}{2mM r^4} + \left( \frac{1}{M^2} + \frac{1}{m^2}\right)\frac{2835}{\pi mM r^5} \right].
     \label{scp}
    \end{split}
\end{eqnarray}
Since there is no tree level contribution in this case, the above is the {\it leading} result in the perturbative expansion. Note that the result is symmetric under the interchange of the two masses, as is expected.\\
\noindent
We now wish to extend Eq.~\ref{scp} for scattering of massive spin-1 and spin-1/2 fields. For $\xi=0$, relevant computations can be seen in~\cite{Holstein:2008sx}. The Feynman diagrams that contribute to this purpose are shown in \ref{triangle2grav}. The solid lines and the thick circles, as earlier, represent the scalar  and the non-minimal vertex. The broken lines will respectively stand for the massive spin-1 (\ref{proca}) and massive spin-1/2 (\ref{dirac}) fields. As in the scalar-scalar scattering, the tree and  (cross-)ladder diagrams do not make any contributions in these cases as well. Note in particular that there are no non-minimal interactions for these spin fields. Hence there are many fewer sub-categories of diagrams here, as compared to the scalar-scalar scattering discussed above.

\section{Massive spin-0-spin-1 interaction }\label{proca}
\begin{figure}[h]
        \begin{minipage}[b]{0.3\linewidth}
            \centering
            \includegraphics[width=0.7\linewidth]{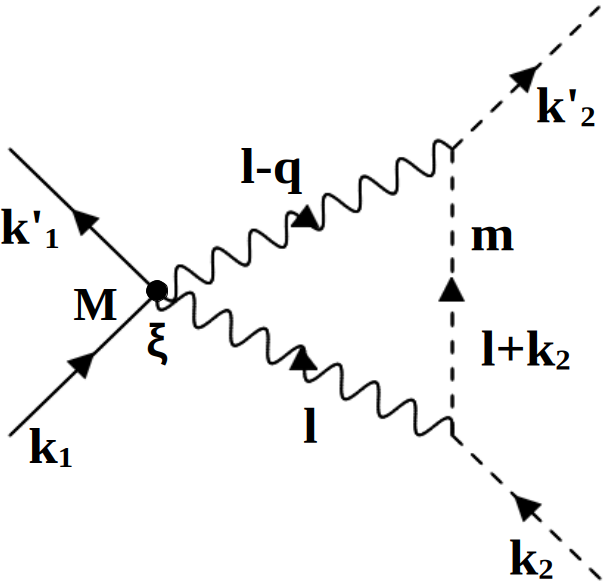}
        \end{minipage}
        \hspace{0.7cm}
        \begin{minipage}[b]{0.3\linewidth}
            \centering
            \includegraphics[width=0.7\linewidth]{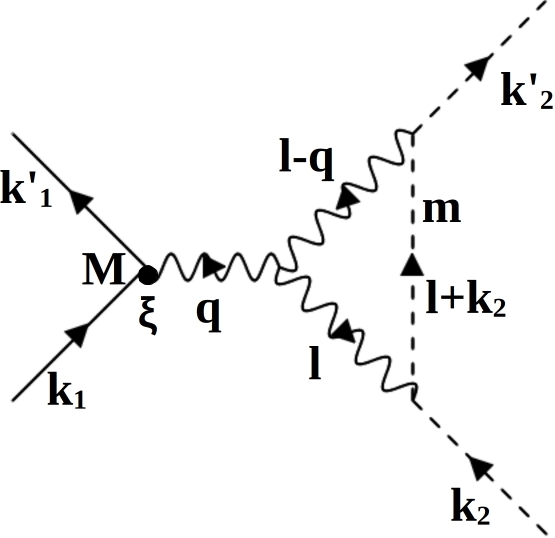}
         \end{minipage} 
        \hspace{0.7cm}
        \begin{minipage}[b]{0.3\linewidth}
            \centering
            \includegraphics[width=0.7\linewidth]{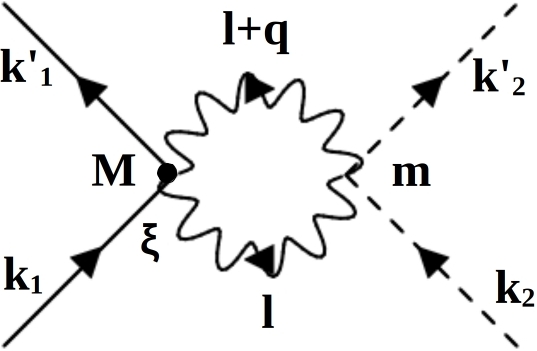}
        \end{minipage} \\
        \\
        \begin{minipage}[b]{0.3\linewidth}
            \centering
            \includegraphics[width=0.9\linewidth]{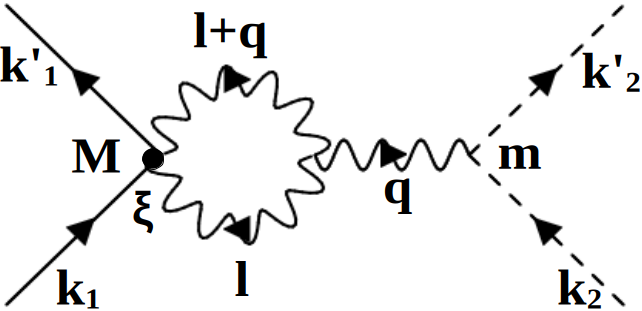}
        \end{minipage}
        \hspace{0.7cm}
        \begin{minipage}[b]{0.3\linewidth}
            \centering
            \includegraphics[width=0.9\linewidth]{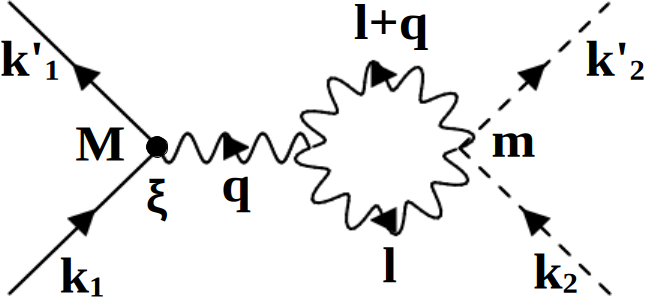}
         \end{minipage} 
        \hspace{0.7cm}
        \begin{minipage}[b]{0.3\linewidth}
            \centering
            \includegraphics[width=0.9\linewidth]{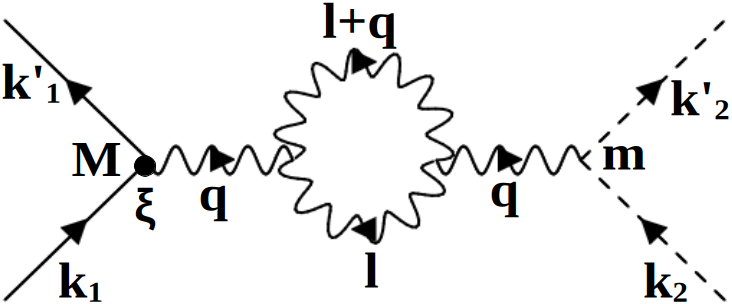}
        \end{minipage}\\
        \caption{\it \small The diagrams for massive spin-0-spin-1 and massive spin-0-spin-1/2 fields scattering. The broken lines  will consecutively represent the massive spin-1 field, and in the next section, a massive spin-1/2 field. We have fixed $(k_1,k_1')$ for the scalar. Since there are no non-minimal interactions for the spin fields, we have much less number of diagrams compared to the scalar-scalar scattering discussed in the preceding section.}
         \label{triangle2grav}
\end{figure}
Let us  compute the long range gravitational potential between a massive spin-0 and a massive spin-1 field at the leading order ${\cal O}(G^2 \xi)$ in this section.
\subsection*{\small a) The triangle diagram :}
There is only one triangle diagram that contributes to our present purpose in this case, given by the first of \ref{triangle2grav}. The Feynman amplitude in the non-relativistic limit reads
\begin{eqnarray}
    \begin{split}
        i \mathcal{M}^{\text{spin-0-spin-1}}_{\text{Triangle}}\big \vert_{\rm NR} =& \int \frac{d^4 l}{(2 \pi)^4} V^{\text{spin-0}\,(2)}_{\rho \sigma \psi \theta \,(\xi)} (l,l-q) \frac{{- i \cal P}^{\psi \theta \mu \nu }}{l^2} \frac{{- i \cal P}^{ \rho \sigma \alpha \beta}}{(l-q)^2}V^{\text{spin-1}\,(1)}_{\tau, \chi, \mu \nu} (k_2,l + k_2,m_v) \epsilon^{\tau}(k_2)\\
        & \times V^ {\text{spin-1}\,(1)}_{\eta, \zeta, \alpha \beta} (l+ k_2,k_2',m_v) \epsilon^{\star \zeta}(k'_2) \frac{-i \mathcal{P}^{\chi \eta}}{[(l+k_2)^2 +m_v^2]} \Big \vert_{\rm NR}\\
 =& \,i\, G^2 \xi \bigg[ \bigg\{ \vec{q}^{\,\,2} \ln \vec{q}^{\,\,2}\Big(\frac{7 \vec{q}^{\,\,6} }{2 m_v^4}  + \frac{\vec{q}^{\,\,4}}{ m_v^2}  - \frac{215 \vec{q}^{\,\,2}}{3}  - 116 m_v^2 \Big) + \frac{35 \pi^2 \vec{q}^{\,\,7}}{32 m_v^3} +\frac{21 \pi^2 \vec{q}^{\,\,5}}{16 m_v}  + 6 \vec{q}^{\,\,4} -\frac{315}{4} m_v \pi^2 \vec{q}^{\,\,3}
        \\& +\frac{8}{3} m_v^2 \vec{q}^{\,\,2} -26 m_v^3 \pi^2 \vec{q}\bigg\} \vec{\epsilon}\cdot\vec{\epsilon}\,\,'  +  \bigg\{ \vec{q}^{\,\,2} \ln \vec{q}^{\,\,2} \Big(-\frac{7 \vec{q}^{\,\,6}}{2 m_v^6}  - \frac{\vec{q}^{\,\,4}}{ m_v^4}    + \frac{215 ^2}{3 m_v^2} \vec{q} + 116 \Big) -\frac{35 \pi^2 \vec{q}^{\,\,7}}{32 m_v^5} 
        \\& -\frac{21 \pi^2 \vec{q}^{\,\,5}}{16 m_v^3}  -\frac{6 \vec{q}^{\,\,4}}{m_v^2}  +\frac{315 \pi^2}{4 m_v} \vec{q}^{\,\,3}-\frac{8 \vec{q}^{\,\,2}}{3} + 26 m_v \pi^2 \vec{q}\bigg\} \vec{k}\cdot\vec{\epsilon} \,\, \vec{k}\cdot\vec{\epsilon}\,\,' +  \bigg\{ \vec{q}^{\,\,2} \ln \vec{q}^{\,\,2}\Big(\frac{7 \vec{q}^{\,\,6}}{8 m_v^6} - \frac{23 \vec{q}^{\,\,4}}{6 m_v^4}
        \\&   - \frac{217  \vec{q}^{\,\,2}}{12 m_v^2} - 120 \Big) - 20 m_v^2 \ln \vec{q}^{\,\,2}+\frac{35 \pi^2 \vec{q}^{\,\,7}}{128 m_v^5}  -\frac{5 \pi^2}{4 m_v^3}- 2 \vec{q}^{\,\,5}+\frac{41 \vec{q}^{\,\,4}}{6 m_v^2} -\frac{137 \pi^2 \vec{q}^{\,\,3}}{8 m_v}  +\frac{4 \vec{q}^{\,\,2}}{3} 
        \\&  +\frac{43}{8} m_v \pi^2 \vec{q}  -\frac{13}{2} m_v \pi^2 \vec{q}+\frac{ 10  m_v^3 \pi^2}{\vec{q}}\bigg\} \vec{q}\cdot\vec{\epsilon} \,\, \vec{q}\cdot\vec{\epsilon}\,\,' +  \bigg\{ \vec{q}^{\,\,2} \ln \vec{q}^{\,\,2} \Big(-\frac{7 \vec{q}^{\,\,6}}{4 m_v^6}  - \frac{\vec{q}^{\,\,4}}{2 m_v^4}  + \frac{215 \vec{q}^{\,\,2}}{6 m_v^2}  + 58 \Big) 
        \\& -\frac{35 \pi^2 \vec{q}^{\,\,7}}{64 m_v^5}  -\frac{21 \pi^2 \vec{q}^{\,\,5}}{32 m_v^3} -\frac{3 \vec{q}^{\,\,4}}{ m_v^2} +\frac{315 \pi^2 \vec{q}^{\,\,3}}{8 m_v} -\frac{4 \vec{q}^{\,\,2}}{3} + 13 m_v \pi^2 \vec{q}\bigg\} i(\vec{k}\times\vec{q})\cdot\vec{S} \bigg],
    \end{split}
    \label{trv}
\end{eqnarray}
where $\e$ and $\vec{S}$ stand respectively  for the polarisation and spin vector of the massive spin-1 field, introduced and discussed in the non-relativistic limit in the centre of mass frame  in Eqs.~\ref{qgs2}, \ref{pol}, \ref{qgs3}, \ref{qgs3'}. Also, $\vec{q}^{\,\,n} = |\vec{q}\,|^{\,n}$ in the above expression and in the following is understood.  We also have abbreviated for convenience 
\begin{eqnarray}
    \mathcal{P}^{\chi \eta} =  \eta^{\chi \eta} + \frac{(l+k_2)^{\chi}(l+k_2)^{\eta}}{m_v^2}. 
\end{eqnarray}
The corresponding  potential reads,
\begin{eqnarray}
    \begin{split}
        V^{\text{spin-0-spin-1}} _{\text{triangle}} (G^2,r, \xi) =& \frac{G^2\xi}{r^4}\Bigg[ \frac{1}{M} \bigg( - 9 m_v^2 +\frac{92 m_v}{\pi r} +\frac{1881}{8 r^2} -\frac{625}{\pi m_v r^3} +\frac{5877}{8 m_v^2 r^4} -\frac{5145}{4\pi m_v^3 r^5} -\frac{26325}{4 m_v^4 r^6} -\frac{227745}{2\pi m_v^5 r^7}  \\
        & +\frac{55125}{4 m_v^6 r^8} -\frac{436590}{\pi m_v^7 r^9} \bigg) \vec{\epsilon}\cdot \vec{\epsilon}\,\,' +\frac{1}{M} \bigg( \frac{13}{2} -\frac{87}{\pi m_v r} -\frac{945}{4 m_v^2 r^2} +\frac{1075}{\pi m_v^3 r^3} -\frac{945}{8 m_v^4 r^4} -\frac{630}{\pi m_v^5 r^5}  \\
        & +\frac{11025}{2 m_v^6 r^6} +\frac{158760}{\pi m_v^7 r^7} \bigg)  \vec{k}\cdot \vec{\epsilon} \,\,\vec{k}\cdot \vec{\epsilon}\,\,' +\frac{1 }{ M r}\bigg( - 13 +\frac{435}{2\pi m_v r} +\frac{2835}{4 m_v^2 r^2} -\frac{7525}{2\pi m_v^3 r^3} +\frac{945}{2 m_v^4 r^4} \\
        & +\frac{2835}{\pi m_v^5 r^5} -\frac{55125}{2 m_v^6 r^6}  -\frac{873180}{\pi m_v^7 r^7} \bigg) (\vec{k}\times \hat r)\cdot \vec{S}+  \frac{1}{M r^2 } \bigg( 10 m_v^2 r^2 +\frac{27}{4} -\frac{3150}{\pi m_v r} -\frac{4932}{m_v^2 r^2}  \\
        & +\frac{68985}{4\pi m_v^3 r^3} +\frac{21375}{2 m_v^4 r^4} -\frac{987525}{2\pi m_v^5 r^5} -\frac{165375}{m_v^6 r^6} +\frac{5675670}{\pi m_v^7 r^7} \bigg)  \hat r\cdot \vec{\epsilon}\,\, \hat r\cdot \vec{\epsilon}\,\,' \bigg].
    \end{split}
    \label{v1}
\end{eqnarray}
\subsection*{\small b) The seagull diagram :}
There is one seagull diagram here given by the second of \ref{triangle2grav}, the Feynman amplitude for which in the non-relativistic limit reads, 
\begin{eqnarray}
    \begin{split}
        i \mathcal{M}^{\text{spin-0-spin-1}}_{\text{seagull}}\big\vert_{\rm NR} =& \int \frac{d^4 l}{(2 \pi)^4} V_{\text{spin-0} \,(1)}^{\rho \sigma \,(\xi)} (q) \frac{-i \cal P_{\rho \sigma \mu \nu }}{q^2} V^{\mu \nu \,(3)  } _{ \alpha \beta \gamma \delta } (l-q,-q) \frac{-i \cal P^{\gamma \delta \psi \theta }}{l^2}  \frac{-i \cal P^{\alpha \beta \lambda \phi}}{(l-q)^2}  V^{\text{spin-1}\,(1)}_{\tau, \chi, \psi \theta} (k_2,l + k_2,m_v) \epsilon^{\tau}(k_2)\\
        & \times V^ {\text{spin-1}\,(1)}_{\eta, \zeta, \lambda \phi} (l+ k_2,k_2',m_v) \epsilon^{\star \zeta}(k'_2) \frac{-i \mathcal{P}^{\chi \eta}}{[(l+k_2)^2 +m_v^2]}\Big\vert_{\rm NR} \\
 =&\,i\, G^2 \xi \bigg[ \bigg\{ \vec{q}^{\,\,2} \ln \vec{q}^{\,\,2}\bigg(
        -\frac{62  \vec{q}^{\,\,6}}{3 m_v^4}
        + \frac{64  \vec{q}^{\,\,4}}{m_v^2}
        + \frac{1828  \vec{q}^{\,\,2}}{3}
        + 696 m_v^2  \bigg)
        -\frac{13 \pi^2  \vec{q}^{\,\,7}}{2 m_v^3}
        -\frac{8  \vec{q}^{\,\,6}}{3 m_v^2} +\frac{209 \pi^2  \vec{q}^{\,\,5}}{8}
        -\frac{68  \vec{q}^{\,\,4}}{3} \\
        &+508 m_v \pi^2  \vec{q}^{\,\,3}
        +112 m_v^3 \pi^2  \vec{q}\bigg\} \vec{\epsilon}\cdot\vec{\epsilon}\,\,'
        +\bigg\{
        \vec{q}^{\,\,2} \ln \vec{q}^{\,\,2}\bigg(
        \frac{62  \vec{q}^{\,\,6}}{3 m_v^6} -\frac{64  \vec{q}^{\,\,4}}{m_v^4}
        -\frac{1828  \vec{q}^{\,\,2}}{3 m_v^2}
        -696   \bigg) +\frac{13 \pi^2  \vec{q}^{\,\,7}}{2 m_v^5}\\
        & +\frac{8  \vec{q}^{\,\,6}}{3 m_v^4} -\frac{209 \pi^2  \vec{q}^{\,\,5}}{8 m_v^3}
        +\frac{68  \vec{q}^{\,\,4}}{3 m_v^2} -\frac{508 \pi^2  \vec{q}^{\,\,3}}{m_v}
        -112 m_v \pi^2  \vec{q}
        \bigg\} \vec{k}\cdot\vec{\epsilon} \,\, \vec{k}\cdot\vec{\epsilon}\,\,' +  \bigg\{
        \ln \vec{q}^{\,\,2}\bigg(
        -\frac{59  \vec{q}^{\,\,8}}{6 m_v^6}
        +\frac{80  \vec{q}^{\,\,6}}{m_v^4}
         \\
        & +\frac{1091  \vec{q}^{\,\,4}}{3 m_v^2} +\frac{1882  \vec{q}^{\,\,2}}{3}
        +16 m_v^2 
        \bigg)
        -\frac{27 \pi^2  \vec{q}^{\,\,7}}{8 m_v^5}
        -\frac{2  \vec{q}^{\,\,6}}{m_v^4}
        +\frac{993 \pi^2  \vec{q}^{\,\,5}}{32 m_v^3}
        -\frac{133  \vec{q}^{\,\,4}}{3 m_v^2}
        +\frac{1125 \pi^2  \vec{q}^{\,\,3}}{4 m_v} -\frac{56  \vec{q}^{\,\,2}}{3} \\
        &
        -\frac{169 m_v \pi^2  \vec{q}}{2}
        +\frac{64 m_v^3 \pi^2 }{\vec{q}}
        \bigg\} \vec{q}\cdot\vec{\epsilon} \,\, \vec{q}\cdot\vec{\epsilon}\,\,'+ \bigg\{
        \ln q^2\bigg(\frac{31   \vec{q}^{\,\,8}}{3 m_v^6}
        -\frac{32   \vec{q}^{\,\,6}}{m_v^4}
        -\frac{914   \vec{q}^{\,\,4}}{3 m_v^2} -348   \vec{q}^{\,\,2}\bigg) +\frac{13  \pi^2  \vec{q}^{\,\,7}}{4 m_v^5}\\
        &  +\frac{4   \vec{q}^{\,\,6}}{3 m_v^4}
        -\frac{209  \pi^2  \vec{q}^{\,\,5}}{16 m_v^3}
        +\frac{34   \vec{q}^{\,\,4}}{3 m_v^2}
        -\frac{254  \pi^2  \vec{q}^{\,\,3}}{m_v}-56  G^2 m_v \pi^2 \xi \vec{q} \bigg\} i(\vec{k}\times\vec{q})\cdot\vec{S} \bigg].
    \end{split}
\end{eqnarray}
The potential is given by,
\begin{eqnarray}
    \begin{split}
        V^{\text{spin-0-spin-1}}_{\text{seagull}}(G^2,r, \xi) =& \frac{G^2 \xi}{M r^4}\bigg[ \bigg(12 m_v^2
        -\frac{526 m}{r}-\frac{3217}{2 \,r^2}+\frac{9140}{\pi m_v \,r^3}-\frac{29655}{4 m_v^2 \, r^4}
        +\frac{38640}{\pi m_v^3 \, r^5} +\frac{110205}{2 m_v^4 \, r^6}+\frac{483840}{\pi m_v^5 \, r^7} \\
        & -\frac{170100}{m_v^6 \, r^8} +\frac{4906440}{\pi m_v^7 \,r^9}\bigg) \vec{\epsilon}\cdot\vec{\epsilon}\,\,'  +\bigg(-28+\frac{522}{\pi m_v \, r}+\frac{1524}{m_v^2 \, r^2}-\frac{9140}{\pi m_v^3 \, r^3}+\frac{9405}{4 m_v^4 \, r^4}  \\
        & -\frac{38640}{\pi m_v^5 \, r^5} -\frac{32760}{m_v^6 \, r^6} -\frac{937440}{\pi m_v^7 \, r^7}\bigg) \vec{k}\cdot\vec{\epsilon} \,\, \vec{k}\cdot\vec{\epsilon}\,\,' + \frac{1}{r} \bigg(56-\frac{1305}{\pi m_v \, r}
        -\frac{4572}{m_v^2 \, r^2}+\frac{31990}{\pi m_v^3 \, r^3}\\
        &  +\frac{9405}{m_v^4 \, r^4}+\frac{173880}{\pi m_v^5 \, r^5} +\frac{163800}{m_v^6 \, r^6}  +\frac{5155920}{\pi m_v^7 \, r^7}\bigg) (\vec{k}\times \hat r)\cdot \vec S
        +\bigg(64 m_v^2+\frac{507}{r^2}
        +\frac{40500}{m_v^2 \, r^4}\\
        & -\frac{223425}{m_v^4 \, r^6} +\frac{4989600}{\pi m_v^5 \, r^7}  +\frac{2041200}{m_v^6 \, r^8} -\frac{63783720}{\pi m_v^7 \, r^9}\bigg) \hat r\cdot\vec{\epsilon} \,\, \hat r\cdot\vec{\epsilon}\,\,' \bigg].
    \end{split}
    \label{v2}
\end{eqnarray}
\subsection*{\small c) The double seagull diagram :}
The Feynman amplitude for the third diagram of \ref{triangle2grav} is,
\begin{eqnarray}
    \begin{split}
         i \mathcal{M}^{\text{spin-0-spin-1}}_{\text{double seagull}}\big\vert_{\rm NR} =& \frac{1}{2!} \int \frac{d^4 l}{(2 \pi)^4} V_{\eta \lambda \rho \sigma \,(\xi)}^{\text{spin-0}\,(2)} (l+q,l) \frac{-i \mathcal{P}^{ \rho \sigma \mu \nu }}{(l+q)^2} \frac{-i \mathcal{P}^{\eta \lambda \rho \sigma}}{l^2}   V^{\text{spin-1\,(2)}}_{\beta,\alpha ,\mu \nu \rho \sigma} (k_2,k_2 ',m_v ) \epsilon^{\beta} \epsilon^{\star  \alpha}\Big\vert_{\rm NR} \\
         &= \,i\,G^2 \xi \ln \vec{q}^{\,\,2} \bigg[\bigg( -\frac{5 \vec{q}^{\,\,4}}{3}  - \frac{208 m_v^2 \vec{q}^{\,\,2}}{3}  \bigg) \vec{\epsilon}\cdot\vec{\epsilon}\,\,' + \bigg( \frac{5 \vec{q}^{\,\,4}}{3 m_v^2}  + \frac{208 \vec{q}^{\,\,2}}{3}  \bigg) \vec{k}\cdot\vec{\epsilon} \,\, \vec{k}\cdot\vec{\epsilon}\,' \\
         &+ \bigg( -\frac{5  \vec{q}^{\,\,4}}{12 m_v^2} - \frac{52 \vec{q}^{\,\,2}}{3}  + 20 m_v^2 \bigg) \vec{q}\cdot\vec{\epsilon} \,\,\vec{q}\cdot\vec{\epsilon}\,' + \bigg( \frac{5 \vec{q}^{\,\,4}}{6 m_v^2}  + \frac{104 \vec{q}^{\,\,2}}{3}  \bigg) i (\vec{k}\times \vec{q})\cdot \vec{S}\bigg]
    \end{split}
\end{eqnarray}
The corresponding potential reads,
\begin{eqnarray}
    \begin{split}
        V^{\text{spin-0-spin-1}} _{\text{double seagull}} (G^2,r, \xi) =& \frac{G^2 \xi}{2 \pi M r^5}\Bigg[ \bigg(188\, m_v
        +\frac{160}{m_v \, r^2}-\frac{175}{m_v^3 \, r^4}
        \bigg) \vec{\epsilon}\cdot\vec{\epsilon}\,\,' +\frac{4}{m_v}\bigg(- 52
        +\frac{25 }{m_v^2 \, r^2}\bigg) \vec{k}\cdot\vec{\epsilon} \,\, \vec{k}\cdot\vec{\epsilon}\,\,'\\
        & +\frac{10}{m_v \, r}
        \bigg(52  -\frac{35}{m_v^2 \, r^2}\bigg) (\vec{k}\times \hat r)\cdot \vec S+\frac{35}{m_v \, r^2}\bigg( - 52 +\frac{45}{m_v^2 \, r^2}\bigg) \hat r\cdot\vec{\epsilon} \,\,\hat r\cdot\vec{\epsilon}\,\,' \bigg].
    \end{split}
    \label{v3}
\end{eqnarray}
\subsection*{\small d) The fish diagrams :}
There are two fish diagrams for this scattering process given by the fourth and fifth of \ref{triangle2grav}. The Feynman amplitudes for them respectively read, 
\begin{eqnarray}
    \begin{split}
        i \mathcal{M}^{\text{spin-0-spin-1}}_{\text{fish-1}}\big\vert_{\rm NR} =& \frac{1}{2!} \int \frac{d^4 l}{(2 \pi)^4} V^{\text{spin-0}\,(2)}_{\rho \sigma \psi \theta \,(\xi)} (l,l+q) \frac{-i \mathcal{P}^{ \rho \sigma \gamma \delta }}{l^2} \frac{-i \mathcal{P}^{ \psi \theta \alpha \beta}}{(l+q)^2} V^{\mu \nu \,(3)} _{\alpha \beta \gamma \delta} (l+q,q) \frac{-i \mathcal{P}_{ \mu \nu\lambda \phi }}{q^2} \\
        & \times V^{\tau, \chi, \lambda \phi}_{\text{spin-1\,(1)}} (k_2,k_2',m_v) \epsilon_{\tau}(k_2) \epsilon^*_{\chi}(k'_2)\Big\vert_{\rm NR} \\
        &= \,i\,G^2 \xi \ln \vec{q}^{\,\,2} \bigg[ \bigg( \frac{310 \vec{q}^{\,\,4}}{3}  + \frac{2060 m_v^2 \vec{q}^{\,\,2}}{3}  \bigg) \vec{\epsilon}\cdot\vec{\epsilon}\,\,' + \bigg( -\frac{310 \vec{q}^{\,\,4}}{3 m_v^2}  - \frac{2060 \vec{q}^{\,\,2}}{3}  \bigg) \vec{k}\cdot\vec{\epsilon} \,\, \vec{k}\cdot\vec{\epsilon}\,'\\
        &\quad+ \bigg( \frac{155 \vec{q}^{\,\,4}}{6 m_v^2}  + \frac{515 \vec{q}^{\,\,2} }{3} \bigg) \vec{q}\cdot\vec{\epsilon} \,\, \vec{q}\cdot\vec{\epsilon}\,' + \bigg( -\frac{155 \vec{q}^{\,\,4}}{3 m_v^2}  - \frac{1030 \vec{q}^{\,\,2} }{3}  \bigg) i (\vec{k}\times \vec{q})\cdot \vec{S}\bigg],
    \end{split}
\end{eqnarray}
and,
\begin{eqnarray}
    \begin{split}
        i \mathcal{M}^{\text{spin-0-spin-1}}_{\text{fish-2}}\big\vert_{\rm NR} =& \frac{1}{2!} \int \frac{d^4 l}{(2 \pi)^4} V_{\text{spin-0}\,(1)}^{\lambda \phi\,(\xi)} (q) \frac{-i \mathcal{P}_{ \lambda \phi \mu \nu }}{q^2}  V^{\mu \nu \,(3)} _{\alpha \beta \gamma \delta} (l,-q) \frac{-i \mathcal{P}^{ \gamma \delta \rho \sigma  }}{(l+q)^2} \frac{-i \mathcal{P}^{ \alpha \beta \psi \theta }}{l^2}  \\\
        & V_{\tau, \chi,\rho \sigma \psi \theta }^{\text{spin-1\,(2)}} (k_2,k_2',m_v) \epsilon^{\tau} (k_2)\epsilon^{\star \chi} (k'_2)\Big\vert_{\rm NR} \\
        &=\,i\, G^2 \xi \ln \vec{q}^{\,\,2} \bigg[\bigg( 16 \vec{q}^{\,\,4} + 752 m^2 \vec{q}^{\,\,2} \bigg) \vec{\epsilon}\cdot\vec{\epsilon}\,\,' + \bigg( -\frac{16 \vec{q}^{\,\,4} }{m^2}  - 752 \vec{q}^{\,\,2} \bigg) \vec{k}\cdot\vec{\epsilon} \,\, \vec{k}\cdot\vec{\epsilon}\,' \\
        &\quad+  \bigg( \frac{4\vec{q}^{\,\,4}}{m^2}  + 188 \vec{q}^{\,\,2} -192 m^2 \bigg) \vec{q}\cdot\vec{\epsilon} \,\,\vec{q}\cdot\vec{\epsilon}\,'+ \bigg( -\frac{8}{m^2} \vec{q}^{\,\,4} - 376 \vec{q}^{\,\,2} \bigg) i (\vec{k}\times \vec{q})\cdot \vec{S} \bigg].
    \end{split}
\end{eqnarray}
Their contribution to the gravitational potential, respectively reads
\begin{equation}
    \begin{split}
        V^{\text{spin-0-spin-1}}_{\text{fish-1}} (G^2,r, \xi) =& \frac{5 G^2 \xi}{2 \pi M r^5}\bigg[ \bigg(- 206 \, m_v
        +\frac{ 725 }{2 m_v \, r^2}+\frac{ 1085 }{m_v^3 \, r^4}
        \bigg)  \vec{\epsilon}\cdot\vec{\epsilon}\,\,' +\frac{2 }{m_v}\bigg(
        103-\frac{310 }{m_v^2 \, r^2}\bigg) \vec{k}\cdot\vec{\epsilon} \,\, \vec{k}\cdot\vec{\epsilon}\,\,' \\
        & + \frac{5}{m_v \, r}
        \bigg(-105 +\frac{ 434}{m_v^2 \,r^2}\bigg) (\vec{k}\times \hat r)\cdot \vec S \bigg] +\frac{5 }{m_v \, r^2}\bigg(
        \frac{721}{2}-\frac{1953}{m_v^2 \, r^2}\bigg) \hat r\cdot\vec{\epsilon} \,\, \hat r\cdot\vec{\epsilon}\,\,' \bigg],\\
        V^{\text{spin-0-spin-1}} _{\text{fish-2}} (G^2,r, \xi) =& \frac{ G^2 \xi}{ \pi M \, r^5}\bigg[   \bigg(- 516 m_v
         -\frac{465n}{m_v r^2}+\frac{420}{m_v^3 r^4}\bigg) \vec{\epsilon}\cdot\vec{\epsilon}\,\,' +\frac{12}{m_v }\bigg(47
        -\frac{20}{m_v^2 \, r^2}\bigg) \vec{k}\cdot\vec{\epsilon} \,\, \vec{k}\cdot\vec{\epsilon}\,\,' \\
        & +\frac{30}{m_v \, r} \bigg(- 47+\frac{28}{m_v^2 r^2}\bigg) (\vec{k}\times \hat r)\cdot \vec S  +\frac{105 }{m_v \, r^2}\bigg( 47-\frac{36}{m_v^2 r^2}\bigg) \hat r\cdot\vec{\epsilon} \,\,\hat r\cdot\vec{\epsilon}\,\,' \bigg].
    \end{split}
    \label{v45}
\end{equation}
\subsection*{\small e) The vacuum polarisation diagram :}
The Feynman amplitude for the last  of \ref{triangle2grav} is given by,
\begin{eqnarray}
    \begin{split}
        \mathcal{M}^{\text{spin-0-spin-1}}_{\text{vac-pol}}\big\vert_{\rm NR} =& V^ {\text{spin-0}\,(1)}_{\mu  \nu \,(\xi)} (q) \dfrac{ - i \mathcal{P}^{\mu \nu \rho \sigma }}{q^2} \Pi_{\rho \sigma \lambda \phi } (q) \ \dfrac{- i \mathcal{P}^{\lambda \phi \gamma \delta}}{q^2}   V_{\beta ,\alpha, \gamma \delta}^{\text{spin-1\,(1)}} (k_2,k_2',m_v) \epsilon^{\beta}(k_2) \epsilon^{\star  \alpha}(k'_2)\Big\vert_{\rm NR} \\
        =& 3 G^2 \xi \, \vec{q}^{\,\,2} \ln \vec{q}^{\,\,2} \bigg[-4
        \left(2 m_v^2 + \vec{q}^{\,\,2}\right) \vec{\epsilon}\cdot\vec{\epsilon}\,\,' +4  \left(2 + \frac{\vec{q}^{\,\,2}}{m_v^2}\right) \vec{k}\cdot\vec{\epsilon} \,\,\vec{k}\cdot\vec{\epsilon}\,\,'  -  \left(2 + \frac{\vec{q}^{\,\,2}}{m_v^2}\right) \vec{q}\cdot\vec{\epsilon} \,\,\vec{q}\cdot\vec{\epsilon}\,\,' \\
        &+2\left(2 + \frac{\vec{q}^{\,\,2}}{m^2}\right) i(\vec{k}\times \vec{q})\cdot \vec S\bigg],
    \end{split}
\end{eqnarray}  
where the expression of the gauge invariant one loop graviton self energy due to itself, $\Pi_{\rho \sigma \lambda \phi } (q)$, can be seen in Eq.~\ref{vpol}~\cite{tHooft:1974toh}. The potential reads,
\begin{eqnarray}
    \begin{split}
        V^{\text{spin-0-spin-1}}_{\text{vac-pol}} (G^2,r, \xi) =& \frac{9G^2 \xi}{\pi M \, r^5}\bigg[  \bigg( 2 m_v
        -\frac{35 }{2 m_v r^2}-\frac{35}{m_v^3 r^4}
        \bigg) \vec{\epsilon}\cdot\vec{\epsilon}\,\,' + \frac{2}{m_v} \bigg(- 1+\frac{10}{m_v^2 r^2}\bigg) \vec{k}\cdot\vec{\epsilon} \,\, \vec{k}\cdot\vec{\epsilon}\,\,'  \\
        & +\frac{5}{m_v r}\bigg( 1 -\frac{14}{m_v^2 r^2}\bigg) (\vec{k}\times \hat r)\cdot \vec S +\frac{35}{m_v r^2}\bigg(-\frac{1}{2 }+\frac{9}{m_v^2 r^2} \bigg)\hat r\cdot\vec{\epsilon} \,\, \hat r\cdot\vec{\epsilon}\,\,' \bigg].
    \end{split}
    \label{v6}
\end{eqnarray}
\subsection{\small The full result :}
Combining now the individual contributions from Eqs.~\ref{v1}, \ref{v2}, \ref{v3}, \ref{v45} and \ref{v6}, we obtain the long range non-minimal gravitational potential between a massive spin-0 and massive spin-1 field at  leading ${\cal O}(G^2 \xi)$,
\begin{eqnarray}
    \begin{split}
        V^{\text{spin-0-spin-1}} (G^2,r, \xi)=& \frac{G^2\xi\,}{M r^4}\bigg[  \bigg( 3 m_v^2  -\frac{1353 m_v }{\pi   r}-\frac{10987  }{8  r^2} \bigg) \vec{\epsilon }\cdot\vec{\epsilon }'  + \bigg(-\frac{43  }{2  }  +\frac{1392 }{\pi  m_v  r} + \frac{5151 }{4 m_v^2  r^2} \bigg) \vec{k}\cdot\vec{\epsilon } \,\, \vec{k}\cdot\vec{\epsilon }' \\
        & + \bigg( \frac{43  }{ r }  -\frac{3505}{\pi  m_v r^2}\bigg) (\vec{k} \times \hat{r})\cdot \vec{S} + \bigg( 74 m_v^2 +\frac{2055  }{4  r^2} \bigg) \hat{r}\cdot\vec{\epsilon } \,\,\hat{r}\cdot\vec{\epsilon }'\bigg] + \ {\cal O}(r^{-7})+ \cdots
    \end{split}
    \label{vp}
\end{eqnarray}

\section{Massive spin-0-spin-1/2 scattering}\label{dirac}
We will next re-compute the diagrams of \ref{triangle2grav}, assuming the broken lines to represent the fermions. Using the expressions given in Eqs.~\ref{fprop} to \ref{qgf5}, we find out the following results.

\subsection*{\small a) The triangle diagram :}

The Feynman amplitude and its non-relativistic limit  for the triangle diagram reads, 
\begin{eqnarray}
    \begin{split}
    i \mathcal{M}^{\text{spin-0-spin-1/2}}_{\text{Triangle}} \big \vert_{\rm NR}=& \int \frac{d^4 l}{(2 \pi)^4} V^{\text{spin-0}\,(2)}_{\rho \sigma \psi \theta \,(\xi)} (l,l-q) \frac{{- i \cal P}^{\psi \theta \mu \nu }}{l^2} \frac{{- i \cal P}^{ \rho \sigma \alpha \beta}}{(l-q)^2}V^{\text{spin-1/2}\,(1)}_{ \mu \nu} (k_2,l + k_2,m_f)\\ 
     &\times V^ {\text{spin-1/2}\,(1)}_{ \alpha \beta} (l+ k_2,k_2',m_f) \bar{u}_{s'}(k_2')  \frac{-i (\gamma ^{\lambda} (l+k_2)_{\lambda} + m_f)}{[(l+k_2)^2 +m_f^2]}  u_s(k_2)\Big \vert_{\rm NR} \\ 
    =&\,i\, G^2 \xi  \bigg[ \bigg\{ \vec{q}^{\,\,2} \ln \vec{q}^{\,\,2} \Big(-\frac{105 \vec{q}^{\,\,4}}{8 m_f^2} + \frac{ 519 \vec{q}^{\,\,2}}{4} + 22 m_f^2\Big)-\frac{315 \pi^2 \vec{q}^{\,\,5}}{64 m_f} +\frac{885 \pi^2 m_f \vec{q}^{\,\,3} }{16}   \\ 
    & - \frac{ 313 \pi^2 m_f^3 \vec{q}}{2} \bigg\} \delta_{ss'}+   \bigg\{ \vec{q}^{\,\,2} \ln \vec{q}^{\,\,2} \left(\frac{105 \vec{q}^{\,\,4}}{16 m_f^4} -\frac{2349 \vec{q}^{\,\,2}}{16 m_f^2} -\frac{3985 m_f^2}{12} \right) \\
    & +\frac{315 \pi^2 \vec{q}^{\,\,5}}{128 m_f^3} -\frac{7283 \pi^2 \vec{q}^{\,\,3}}{128 m_f} -\frac{179 \pi^2 m_f \vec{q} }{4} \bigg\} i(\vec{k}\times\vec{q})\cdot\vec{S}_{1/2}  \bigg].
    \end{split}
\end{eqnarray}
The corresponding potential is given by,
\begin{equation}
    \begin{split}
        V^{\text{spin-0-spin-1/2}} _{\text{triangle}} (G^2,r, \xi) =& \frac{G^2\xi}{M r^4}\bigg[\bigg(-\frac{313 m_f^2}{8}-\frac{33 m_f}{2 \pi  r}- \frac{2655}{16  r^2}
        +\frac{7785}{4\pi m_f  r^3} -\frac{14175}{32 m_f^2  r^4} - \frac{33075}{4 \pi m_f^3  r^5}\bigg) \delta_{ss'}\\
        &+\frac{1}{r}\bigg( 45 - \frac{19925}{16\pi m_f r} - \frac{65547}{64 m_f^2  r^2} + \frac{246645}{16 \pi m_f^3  r^3} - \frac{14175}{8 m_f^4  r^4}  - \frac{297675}{8 \pi m_f^5  r^5}\bigg) (\vec{k}\times \hat r)\cdot \vec S_{1/2}\bigg]. \\
    \end{split}
    \label{fpot1}
\end{equation}
\subsection*{\small b) The seagull diagram :}
The Feynman amplitude for the second diagram of \ref{triangle2grav} reads  in the non-relativistic limit, 
\begin{eqnarray}
    \begin{split}
        i \mathcal{M}^{\text{spin-0-spin-1/2}}_{\text{seagull}}\big \vert_{\rm NR} =& \int \frac{d^4 l}{(2 \pi)^4} V_{\text{spin-0} \,(1)}^{\rho \sigma \,(\xi)} (q) \frac{-i \cal P_{\rho \sigma \mu \nu }}{q^2} V^{\mu \nu \,(3)  } _{ \alpha \beta \gamma \delta } (l-q,-q) \frac{-i \cal P^{\gamma \delta \psi \theta }}{l^2}  \frac{-i \cal P^{\alpha \beta \lambda \phi}}{(l-q)^2}  V^{\text{spin-1/2}\,(1)}_{\tau, \chi, \psi \theta} (k_2,l + k_2,m_f) \\
        &  \times  V^ {\text{spin-1/2}\,(1)}_{ \lambda \phi} (l+ k_2,k_2',m_f)  \bar{u}_{s'}(k_2')  \frac{-i (\gamma ^{\lambda} (l+k_2)_{\lambda} + m_f)}{[(l+k_2)^2 +m_f^2]}  u_s(k_2)\Big \vert_{\rm NR} \\
        =& \,i\,G^2 \xi\bigg[ \bigg\{\vec{q}^{\,\,2} \ln \vec{q}^{\,\,2}\bigg( \frac{8  \vec{q}^{\,\,4} }{m_f^2} - 1085  \vec{q}^{\,\,2} - 1770 m_f^2 \bigg) + \frac{89 \pi ^2  \vec{q}^{\,\,5}}{ 32 m_f}- \frac{2515 \pi ^2 m_f  \vec{q}^{\,\,3} }{4}\\
        &  -406 \pi ^2 m_f^3  \vec{q} \bigg\} \delta _{ss'} + \bigg\{ \vec{q}^{\,\,2} \ln \vec{q}^{\,\,2}\bigg( - \frac{4   \vec{q}^{\,\,4}}{ m_f^4} +\frac{3475   \vec{q}^{\,\,2}}{4 m_f^2} + 230   \bigg) - \frac{89 \pi ^2   \vec{q}^{\,\,5}}{64m_f^3} +\frac{7163 \pi ^2   \vec{q}^{\,\,3}}{16 m_f} \\
        & -1215 \pi ^2  m_f  \vec{q}  \bigg\} i (\vec{k}\times \vec{q})\cdot \vec{S}_{1/2} \bigg].
    \end{split}
\end{eqnarray}
The corresponding potential is found to be,
\begin{equation}
    \begin{split}
        V^{\text{spin-0-spin-1/2}}_{\text{seagull}}(G^2,r, \xi) = &\frac{G^2\xi}{M r^4}\bigg[  \bigg(- \frac{203 m_f^2}{2} + \frac{2655 m_f}{2\pi r} + \frac{7545}{ 4 r^2} - \frac{16275}{\pi m_f r^3} + \frac{4005}{16 m_f^2 r^4}
        + \frac{5040}{\pi m_f^3 r^5}\bigg) \,\delta_{ss'}\, \\
        & +\frac{1}{ r}\bigg( 1215  + \frac{ 1725 }{2 \pi m_f r} + \frac{ 64467}{8 m_f^2 r^2} - \frac{364875}{4\pi m_f^3 r^3} + \frac{4005}{4 m_f^4 r^4} + \frac{22680}{\pi m_f^5 r^5}\bigg) (\vec{k}\times \hat r)\cdot \vec S_{1/2} \bigg].
    \end{split}
\label{fpot2}
\end{equation}
\subsection*{\small c) The double seagull diagram :}
The Feynman amplitude and its non-relativistic limit for the third diagram of \ref{triangle2grav} reads, 
\begin{eqnarray}
    \begin{split}
          i \mathcal{M}^{\text{spin-0-spin-1/2}}_{\text{double seagull}}\big \vert_{\rm NR} =& \frac{1}{2!} \int \frac{d^4 l}{(2 \pi)^4} V_{\eta \lambda \rho \sigma \,(\xi)}^{\text{spin-0}\,(2)} (l+q,l) \frac{-i \mathcal{P}^{ \rho \sigma \mu \nu }}{(l+q)^2} \frac{-i \mathcal{P}^{\eta \lambda \rho \sigma}}{l^2}   \bar{u}_{s'}(k_2')   V^{\text{spin-1/2\,(2)}}_{\mu \nu \rho \sigma} (k_2,k_2 ',m_f) u_s(k_2)\Big \vert_{\rm NR}  \\ 
        =& \,\frac{2924 i}{3} G^2 m_f^2 \xi  \vec{q}^{\,\,2} \ln \vec{q}^{\,\,2} \delta _{ss'} + \,\frac{214 }{3} G^2 i \xi  \vec{q}^{\,\,2} \ln \vec{q}^{\,\,2} (\vec{k}\times \vec{q})\cdot \vec{S}_{1/2}.
\label{}
    \end{split}
\end{eqnarray}
Accordingly, the potential reads,
\begin{eqnarray}
    \begin{split}
        V^{\text{spin-0-spin-1/2}} _{\text{double seagull}} (G^2,r, \xi) =& -\frac{731 G^2 m_f \xi  \delta _{ss'}}{ \pi  M r^5} + \frac{535 G^2 \xi  (\vec{k} \times \hat{r})\cdot \vec{S}_{1/2}}{2 \pi  m M r^6}.
    \end{split}
    \label{fpot3}
\end{eqnarray}
\subsection*{\small d) The fish diagrams :}
The Feynman amplitudes for the fourth and fifth diagrams of \ref{triangle2grav} in the non-relativistic limit are respectively given by, 
\begin{eqnarray}
    \begin{split}
          i \mathcal{M}^{\text{spin-0-spin-1/2}}_{\text{fish-1}} \big\vert_{\rm NR} =& \frac{1}{2!} \int \frac{d^4 l}{(2 \pi)^4} V^{\text{spin-0}\,(2)}_{\rho \sigma \psi \theta \,(\xi)} (l,l+q) \frac{-i \mathcal{P}^{ \rho \sigma \gamma \delta }}{l^2} \frac{-i \mathcal{P}^{ \psi \theta \alpha \beta}}{(l+q)^2} V^{\mu \nu \,(3)} _{\alpha \beta \gamma \delta} (l+q,q) \frac{-i \mathcal{P}_{ \mu \nu\lambda \phi }}{q^2} \\
        & \times \bar{u}_{s'}(k_2') V^{ \lambda \phi}_{\text{spin-1/2\,(1)}} (k_2,k_2',m_f) u_s(k_2) \Big\vert_{\rm NR} \\ 
        =& \frac{17930 i}{3} G^2 m_f^2 \xi  \vec{q}^{\,\,2} \ln \vec{q}^{\,\,2} \delta _{ss'}-\,\frac{515 i}{3} G^2 i \xi  \vec{q}^{\,\,2} \ln \vec{q}^{\,\,2} (\vec{k}\times \vec{q})\cdot\vec{S}_{1/2},
    \end{split}
\end{eqnarray}
and
\begin{eqnarray}
    \begin{split}
        i \mathcal{M}^{\text{spin-0-spin-1/2}}_{\text{fish-2}}\big\vert_{\rm NR} =& \frac{1}{2!} \int \frac{d^4 l}{(2 \pi)^4} V_{\text{spin-0}\,(1)}^{\lambda \phi\,(\xi)} (q) \frac{-i \mathcal{P}_{ \lambda \phi \mu \nu }}{q^2}  V^{\mu \nu \,(3)} _{\alpha \beta \gamma \delta} (l,-q) \frac{-i \mathcal{P}^{ \gamma \delta \rho \sigma  }}{(l+q)^2} \frac{-i \mathcal{P}^{ \alpha \beta \psi \theta }}{l^2} \\
        & \times \bar{u}_{s'}(k_2') V_{\rho \sigma \psi \theta }^{\text{spin-1/2\,(2)}} (k_2,k_2',m_f) u_s(k_2) \Big\vert_{\rm NR} \\ 
        =& -\, i 5600 G^2 m_f^2 \xi  \vec{q}^{\,\,2} \ln \vec{q}^{\,\,2} \delta _{ss'} -\,i 400 G^2 i \xi  \vec{q}^{\,\,2} \ln \vec{q}^{\,\,2} (\vec{k}\times \vec{q})\cdot \vec{S}_{1/2}.
    \end{split}
\end{eqnarray}
The corresponding two body potentials read, 
\begin{equation}
    \begin{split}
        V^{\text{spin-0-spin-1/2}}_{\text{fish-1}} (G^2,r, \xi) =& -\frac{8965 G^2 m_f \xi  \delta _{ss'}}{2 \pi  M r^5} - \frac{2575 G^2 \xi  (\vec{k} \times \hat{r})\cdot\vec{S}_{1/2}}{4 \pi  m_f M r^6}, \\
        V^{\text{spin-0-spin-1/2}} _{\text{fish-2}} (G^2,r, \xi) =& \frac{4200 G^2 m_f \xi  \delta _{ss'}}{ \pi  M r^5}-\frac{1500 G^2 \xi  (\vec{k} \times \hat{r})\cdot \vec{S}_{1/2}}{ \pi  m_f M r^6}.
    \end{split}
    \label{fpot4}
\end{equation}
\subsection*{\small e) The vacuum polarisation diagram :}
The Feynman amplitude for the last diagram of \ref{triangle2grav} reads in the non-relativistic limit,
\begin{eqnarray}
    \begin{split}
        i\mathcal{M}^{1/2}_{\text{vac-pol}}\big\vert_{\rm NR} =& V^ {\text{spin-0}\,(1)}_{\mu  \nu \,(\xi)} (q) \dfrac{ - i \mathcal{P}^{\mu \nu \rho \sigma }}{q^2} \Pi_{\rho \sigma \lambda \phi } (q) \ \dfrac{- i \mathcal{P}^{\lambda \phi \gamma \delta}}{q^2}  \bar{u}_{s'}(k_2') V_{ \gamma \delta}^{\text{spin-1/2\,(1)}} (k_2,k_2',m_f) u_s(k_2) \Big\vert_{\rm NR}\\
        =& -276i G^2 m^2 \xi  \vec{q}^{\,\,2} \ln \vec{q}^{\,\,2} \delta _{ss'} + 6 G^2 \xi  i \vec{q}^{\,\,2} \ln \vec{q}^{\,\,2} (\vec{k}\times \vec{q})\cdot \vec{S}_{1/2},
    \end{split}
\end{eqnarray} 
where the expression of $\Pi_{\rho \sigma \lambda \phi } (q)$ can be seen in Eq.~\ref{vpol}. The corresponding potential reads,
\begin{eqnarray}
    \begin{split}
        V^{\text{spin-0-spin-1/2}}_{\text{vac-pol}} (G^2,r, \xi) =& \frac{207 G^2 m_f \xi  \delta _{ss'}}{\pi  M r^5} +\frac{45 G^2 \xi  (\vec{k} \times \hat{r})\cdot \vec{S}_{1/2}}{2 \pi  m_f M r^6}. 
    \end{split}
    \label{fpot5}
\end{eqnarray}
\subsection{\small The full result :}
Combining now the individual contributions of Eqs.~\ref{fpot1}, \ref{fpot2}, \ref{fpot3}, \ref{fpot4}, \ref{fpot5}, we obtain the long range non-minimal  gravitational potential between a massive scalar and fermion at the leading order ${\cal O}(G^2\xi)$,
\begin{eqnarray}
    \begin{split}
        V^{\text{spin-0-spin-1/2}} (G^2,r, \xi, ss')=& \frac{G^2\xi}{M r^4}\bigg[ \bigg( -\frac{1125 m_f^2  }{8 }+\frac{1009 m_f  }{2 \pi   r}+\frac{27525 }{16 r^2} \bigg) \delta _{ss'} + \bigg( \frac{1260 }{r} -\frac{35785 }{16 \pi  m_f  r^2}\bigg) (\vec{k} \times \hat{r})\cdot \vec{S}_{ss'} \bigg] \\ & +  \ {\cal O}(r^{-7})+\cdots.
    \end{split}
    \label{fp}
\end{eqnarray}
%
\section{Discussion}\label{disc}
We have computed the effect of gravity-scalar $\xi R\phi^2$ non-minimal coupling in the two body long range gravitational potential for massive fields. We have considered the 2-2 scattering between  scalars,  scalar-spin-1, and scalar-spin-1/2 fields. From the non-relativistic limit of the scattering Feynman amplitudes, we have computed the gravitational   potentials in Eqs.~\ref{scp}, \ref{vp} and \ref{fp}. These are the main results of this paper. Note that there is no tree level contribution here, and hence the ${\cal O}(G^2 \xi)$ results found here are leading.  Also, the leading behaviour of the potential is $\sim r^{-4}$. Let us now compare our result with the well known $\xi=0$ case~\cite{Bjerrum-Bohr:2002gqz},
\begin{equation}
    V(r, \xi=0)= -\frac{GMm}{r}\left( 1 + \frac{3G(M+m)}{r} +  \frac{41 G \hbar}{10 \pi r^2}\right).
\end{equation}
The higher radial powers of our case compared to the above result certainly originate from  the explicit appearance of the transfer momentum carried by the graviton lines in the non-minimal vertices of various scattering amplitudes (cf., discussion below Eq.~\ref{nmv1}). 

Imagine now that $M \gg m$ and we compare the leading $r^{-4}$ part of Eq.~\ref{scp} with that of $V(r, \xi=0)$.  The ratio of this leading part and the most subleading (quantum) part of $V(r, \xi=0)$ reads ($\hbar=1$),
$$3.06 \xi \times \frac{M}{m}\times \frac{1}{mr} .$$
The masses appearing above are to be understood as their inverse Compton wavelengths. Let us now imagine a particle of electron mass on the surface of the Earth. We take $M \sim 10^{24}$kg, $m \sim 10^{-31}$kg, $r \sim 10^{6}$m (the radius of the earth), and taking the Compton wavelength of $m$ to be $\sim 10^{-12}$m, we see that the above ratio is about $10^{37}\xi$. Thus, for any reasonable value less than unity of the non-minimal coupling parameter, Eq.~\ref{scp} will dominate over the $r^{-3}$ part of the one loop minimal gravitational potential.  The second term of $V(r, \xi=0)$ is, however, much more dominant. Nevertheless, it is clear that the effect of this $r^{-4}$ term on the test particles near a massive object like a black hole can be interesting, for maybe in future observation it can distinguish the non-minimal interaction, if it really exists. 

For a massive vector or spinor  field, however, the ratio becomes
$$\sim \frac{m}{M}\times \frac{1}{Mr}$$
which is much subleading if we take the earlier values of the parameters.

 Note also that apart from having different coefficients, the scalar-scalar potential Eq.~\ref{scp} is symmetric under the interchange of the two masses, in contrast to the result for scalar-vector and scalar-fermion  interactions, Eqs.~\ref{vp} and \ref{fp}. This is expected, as for the scalar-scalar interaction the particles are of the same species (with the same taken value of the non-minimal coupling, $\xi$). Hence, interchanging the masses leaves the potential unchanged. For the two other interactions, however, first, the particle species are different with  different kinds of propagators, and second, there is no non-minimal interaction for the non-spin-0 fields. This (a)symmetry feature  is also  evident from the various Feynman diagrams.

It seems to be an important task to understand the effect of motion or macroscopic spin of the massive scalar body on the gravitational potential, in the context of $\xi R \phi^2$ interaction.
It will be further important to understand the effect of $\xi$ on gravitational light bending. Investigation of various other scalar tensor theories~\cite{Clifton:2011jh} seems also to be an important task. We hope to come back to these issues in our future publications.

\bigskip
\bigskip
\bigskip
\noindent
{\bf \large Acknowledgements :} AKN's research is supported by the research fellowship of the University Grants Commission, Govt. of India (NTA Ref. No./Student ID : 221610099618). ASM's research is supported by the fellowship of the State Government Departmental Fellowship Scheme, Govt. of West Bengal (India). The authors would like to thank anonymous referee for useful comments on an earlier version of this manuscript.

\bigskip
\appendix
\labelformat{section}{Appendix #1} 
\section{Some useful formulae}\label{A}

\subsection*{\small a) List of Fourier transforms : }

The 3-D Fourier transform of a function $f(\vec{q})$ is defined as 
\begin{equation*}
     \int \dfrac{d^3 \vec{q}}{(2\pi)^3}\, e^{i \vec{q}\cdot \vec{r}} \, f(\,\vec{q}\,) \, = \, g(\,\vec{r}\,)
 \end{equation*}
 \begin{table}[h]
\centering
\caption{\small \it Table for required Fourier transforms}
\renewcommand{\arraystretch}{2.5}
\begin{tabular}{||c|c|c||c|c|c||}
\hline 
{\bf sl.no.} & $\bm{f(\,\vec{q}\,)}$ & $\bm{g(\,\vec{r}}\,)$ & {\bf sl.no.} & $\bm{f(\,\vec{q}}\,)$ & $\bm{g(\,\vec{r}}\,)$ \\ [6pt]
\hline 
$ 1. $ & $(\,\vec{q}\,)^n \, \ln \vec{q}^{\,2}$ & $ \,\dfrac{A(n)}{r^{n+3}}$ & $ 2. $ & $(\,\vec{q}\,)^{2n} \, $ & $\big(-{\vec{\p}}^2\big)^n \, \delta^{(3)}(\vec{r})$  \\ [6pt]
\hline 
$3. $ & $\,q_i\,(\,\vec{q}\,)^n \,\ln \vec{q}^{\,2}$ & $A(n) \dfrac{i (n+3) r_i}{r^{n+5}}$ & $ 4. $ & $\,q_i \,(\,\vec{q}\,)^{2n}   $ & $-i\,\partial_i\big(-{\vec{\p}}^2\big)^n \, \delta^{(3)}(\vec{r})$ \\ [6pt]
\hline 
$5. $ & $\,q_i\,q_j \,(\,\vec{q}\,)^n \,\ln \vec{q}^{\,2}$ & $A(n) (n+3)\Bigg(\,\dfrac{\delta_{ij}}{r^{n+5}}-(n+5)\,\dfrac{r_i \, r_j}{r^{n+7}}\Bigg)$ &  $ 6. $ & $(\,\vec{q}\,)^{k} $ $ $ & $ \dfrac{B(k)}{r^{k+3}}$ \\ [6pt]
\hline
$ 7. $ & $\,q_i\,q_j \,(\,\vec{q}\,)^k $ & $B(k) (k+3)\Bigg(\,\dfrac{\delta_{ij}}{r^{k+5}}-(k+5)\,\dfrac{r_i \, r_j}{r^{k+7}}\Bigg)$& $8. $ & $\,q_i\,(\,\vec{q}\,)^k  $ & $B(k) \dfrac{i (k+3) r_i}{r^{k+5}}$  \\ [6pt]
\hline
\end{tabular}
\end{table}\\
where $n$ is any positive integer, and $k$ is any odd positive integer. $A(n)$ and $B(k)$ are defined as,
        \begin{eqnarray*}
            A(n) = \dfrac{\big(-1\big)^{[n/2+1]} \,\Gamma(n)}{2\,\pi^{\,1/2\big(3-(-1)^n\big)}}, \, \qquad 
            B(k) = \dfrac{2^k \, \Gamma\left( \dfrac{3+k}{2}\right)}{\pi^{\,3/2}\,\Gamma\left( -\dfrac{k}{2}\right) },  
           \end{eqnarray*}  
where the $[.]$ appearing in $A(n)$ is the greatest integer function, i.e., it returns the largest integer less than or equal to the given number.

\subsection*{\small b) List of some essential integrals :}
\begin{eqnarray}
 &&       \int \dfrac{d^d l}{(2 \pi)^d} \dfrac{1}{l^2 (l+q)^2} =  -   \dfrac{i }{16 \pi^2} \ln q^2, \qquad   \int \dfrac{d^d l}{(2 \pi)^d} \dfrac{l_{\mu}}{l^2 (l+q)^2} = \dfrac{i q_{\mu}}{32 \pi^2} \ln q^2 \nonumber\\ &&
        \int \dfrac{d^d l}{(2 \pi)^d} \dfrac{l_{\mu}l_{\nu}}{l^2 (l+q)^2} = - \dfrac{i \ln q^2}{64 \pi^2} \Big[\dfrac{4}{3} q_{\mu} q_{\nu} - \dfrac{1}{3} q^2 \eta_{\mu \nu}\Big]\nonumber\\ &&
        \int \dfrac{d^d l}{(2 \pi)^d} \dfrac{1}{l^2 (l+q)^2 ((l+k)^2 - m^2)} = -\dfrac{i}{32 \pi^2  m^2} \Bigg[ \ln q^2 + \dfrac{\pi^2m}{q}  \Bigg].
\end{eqnarray}
\begin{eqnarray}
        \int \dfrac{d^d l}{(2 \pi)^d} \dfrac{l_{\mu}}{l^2 (l+q)^2 ((l+k)^2 - m^2)} = \dfrac{i}{32 \pi^2 m^2} \left[ \left\{\left(-1 -  \dfrac{q^2}{2m^2}\right)  \ln q^2 - \dfrac{1}{4} \dfrac{q^2}{m^2}  \dfrac{\pi^2 m}{q} \right\}  k_{\mu} + \left\{ \ln q^2 + \dfrac{\pi^2 m}{2 q} \right\}q_{\mu}  \right].
\end{eqnarray}
\begin{eqnarray}
   &&     \int \dfrac{d^d l}{(2 \pi)^d} \dfrac{l_{\mu} l_{\nu}}{l^2 (l+q)^2 ((l+k)^2 - m^2)} = \dfrac{i}{32 \pi^2 m^2} \left[ q_{\mu} q_{\nu} \left( - \ln q^2 -\dfrac{3}{8} \dfrac{\pi^2 m}{q}  \right) - \frac{k_{\mu} k_{\nu} q^2}{2m^2} \left( \ln q^2 +  \dfrac{\pi^2  q^2}{4m q}   \right) \right. \nonumber \\ && \left.
        + \dfrac{1}{2}(q_{\mu} k_{\nu} + q_{\nu} k_{\mu}) \left\{ \left(  1+  \dfrac{q^2}{m^2}  \right)\ln q^2 + \dfrac{3 \pi^2 m q}{8}   \right\} + \dfrac{q^2}{4} \eta_{\mu \nu} \left(  \ln q^2 +   \dfrac{\pi^2 m}{2 q}  \right)  \right].
\end{eqnarray}
\begin{eqnarray}
    \begin{split}
        \int\frac{d^4l}{(2\pi)^4} \frac{l_\mu l_\nu l_\alpha}{l^2(l+q)^2((l+k)^2-m^2)} =& \frac{i}{32\pi^2m^2}\Bigg[ q_\mu q_\nu q_\alpha\Bigg( \ln q^2+\frac5{16}\dfrac{\pi^2 m}{q} \Bigg)+k_\mu k_\nu k_\alpha\Bigg(-\frac16 \frac{q^2}{m^2}\Bigg)\\
        & + \big(q_\mu k_\nu k_\alpha + q_\nu k_\mu k_\alpha + q_\alpha k_\mu k_\nu\big)\Bigg(\frac13\frac{q^2}{m^2} \ln q^2+ \frac1{16}\frac{q^2}{m^2}\dfrac{\pi^2 m}{q} \Bigg) \\
        & + \big(q_\mu q_\nu k_\alpha + q_\mu q_\alpha k_\nu + q_\nu q_\alpha k_\mu \big)\Bigg(\left(-\frac13 - \frac12\frac{q^2}{m^2}\right) \ln q^2 -\frac{5}{32}\frac{q^2}{m^2}\dfrac{\pi^2 m}{q} \Bigg) \\ 
        & + \big(\eta_{\mu\nu}k_\alpha + \eta_{\mu\alpha}k_\nu + \eta_{\nu\alpha}k_\mu\big)\times \frac{q^2}{12} \ln q^2  \\ 
        &+ \big(\eta_{\mu\nu}q_\alpha + \eta_{\mu\alpha}q_\nu + \eta_{\nu\alpha}q_\mu\big)\left(-\frac16q^2 \ln q^2 -\frac1{16}q^2\dfrac{\pi^2 m}{q} \right) \Bigg].
    \end{split}
\end{eqnarray}

\end{document}